# The putative center in NGC 1052

Anne-Kathrin Baczko[1,2,⋆], Matthias Kadler[3], Eduardo Ros[2], Christian M. Fromm[3,4,2], Maciek Wielgus[2], Manel Perucho[5,6], Thomas P. Krichbaum[2], Mislav Baloković[7], Lindy Blackburn[8,9], Chi-kwan Chan[10,11,12], Sara Issaoun[9,⋆⋆], Michael Janssen[13,2], Luca Ricci[3,2], Kazunori Akiyama[14,15,8], Ezequiel Albentosa-Ruíz[5], Antxon Alberdi[16], Walter Alef[2], Juan Carlos Algaba[17], Richard Anantua[18,142,8,9], Keiichi Asada[19], Rebecca Azulay[5,6,2], Uwe Bach[2], David Ball[10], Bidisha Bandyopadhyay[20], John Barrett[14], Michi Bauböck[21], Bradford A. Benson[22,23], Dan Bintley[24,25], Raymond Blundell[9], Katherine L. Bouman[26], Geoffrey C. Bower[27,28], Hope Boyce[29,30], Michael Bremer[31], Christiaan D. Brinkerink[13], Roger Brissenden[8,9], Silke Britzen[2], Avery E. Broderick[32,33,34], Dominique Broguiere[31], Thomas Bronzwaer[13], Sandra Bustamante[35], Do-Young Byun[36,37], John E. Carlstrom[38,23,39,40], Chiara Ceccobello[41], Andrew Chael[42], Dominic O. Chang[8,9], Koushik Chatterjee[8,9], Shami Chatterjee[43], Ming-Tang Chen[27], Yongjun Chen[44,45], Xiaopeng Cheng[36], Ilje Cho[16,36,46], Pierre Christian[47], Nicholas S. Conroy[48,9], John E. Conway[41], James M. Cordes[43], Thomas M. Crawford[23,38], Geoffrey B. Crew[14], Alejandro Cruz-Osorio[49,4], Yuzhu Cui[50,51], Rohan Dahale[16], Jordy Davelaar[114,⋆⋆], Mariafelicia De Laurentis[54,4,55], Roger Deane[56,57,58], Jessica Dempsey[24,25,59], Gregory Desvignes[2,60], Jason Dexter[61], Vedant Dhruv[21], Indu K. Dihingia[51], Sheperd S. Doeleman[8,9], Sean Taylor Dougall[10], Sergio A. Dzib[31,2], Ralph P. Eatough[62,2], Razieh Emami[9], Heino Falcke[13], Joseph Farah[63,64], Vincent L. Fish[14], Edward Fomalont[65], H. Alyson Ford[10], Marianna Foschi[16], Raquel Fraga-Encinas[13], William T. Freeman[66,67], Per Friberg[24,25], Antonio Fuentes[16], Peter Galison[8,68,69], Charles F. Gammie[21,48,70], Roberto García[31], Olivier Gentaz[31], Boris Georgiev[10], Ciriaco Goddi[71,72,73,74], Roman Gold[75], Arturo I. Gómez-Ruiz[76,77], José L. Gómez[16], Minfeng Gu[44,78], Mark Gurwell[9], Kazuhiro Hada[79,80], Daryl Haggard[29,81], Kari Haworth[9], Michael H. Hecht[14], Ronald Hesper[82], Dirk Heumann[10], Luis C. Ho[83,84], Paul Ho[19,25,24], Mareki Honma[85,86,87], Chih-Wei L. Huang[19], Lei Huang[44,78], David H. Hughes[76], C. M. Violette Impellizzeri[88,65], Makoto Inoue[19], David J. James[89,90], Buell T. Jannuzi[10], Britton Jeter[19], Wu Jiang[44], Alejandra Jiménez-Rosales[13], Michael D. Johnson[8,9], Svetlana Jorstad[91], Abhishek V. Joshi[21], Taehyun Jung[36,37], Mansour Karami[32,33], Ramesh Karuppusamy[2], Tomohisa Kawashima[92], Garrett K. Keating[9], Mark Kettenis[93], Dong-Jin Kim[2], Jae-Young Kim[94,2], Jongsoo Kim[36], Junhan Kim[95], Motoki Kino[15,96], Jun Yi Koay[19], Prashant Kocherlakota[4], Yutaro Kofuji[85,87], Shoko Koyama[97,19], Carsten Kramer[31], Joana A. Kramer[2], Michael Kramer[2], Cheng-Yu Kuo[98,19], Noemi La Bella[13], Tod R. Lauer[99], Daeyoung Lee[21], Sang-Sung Lee[36], Po Kin Leung[100], Aviad Levis[26], Zhiyuan Li[101,102], Rocco Lico[103,16], Greg Lindahl[9], Michael Lindqvist[41], Mikhail Lisakov[104], Jun Liu[2], Kuo Liu[2], Elisabetta Liuzzo[105], Wen-Ping Lo[19,106], Andrei P. Lobanov[2], Laurent Loinard[107], Colin J. Lonsdale[14], Amy E. Lowitz[10], Ru-Sen Lu[44,108,2], Nicholas R. MacDonald[2], Jirong Mao[109,110,111], Nicola Marchili[105,2], Sera Markoff[112,113], Daniel P. Marrone[10], Alan P. Marscher[91], Iván Martí-Vidal[5,6], Satoki Matsushita[19], Lynn D. Matthews[14], Lia Medeiros[114,⋆⋆], Karl M. Menten[2], Daniel Michalik[115,116], Izumi Mizuno[24,25], Yosuke Mizuno[51,117,4], James M. Moran[8,9], Kotaro Moriyama[4,14,85], Monika Moscibrodzka[13], Wanga Mulaudzi[112], Cornelia Müller[2,13], Hendrik Müller[2], Alejandro Mus[72,105], Gibwa Musoke[112,13], Ioannis Myserlis[118], Andrew Nadolski[48], Hiroshi Nagai[15,86], Neil M. Nagar[20], Dhanya G. Nair[20], Masanori Nakamura[119,19], Gopal Narayanan[35], Iniyan Natarajan[9,8], Antonios Nathanail[120,4], Santiago Navarro Fuentes[118], Joey Neilsen[121], Roberto Neri[31], Chunchong Ni[33,34,32], Aristeidis Noutsos[2], Michael A. Nowak[122], Junghwan Oh[93], Hiroki Okino[85,87], Héctor Raúl Olivares Sánchez[123], Gisela N. Ortiz-León[76,2], Tomoaki Oyama[85], Feryal Özel[124], Daniel C. M. Palumbo[8,9], Georgios Filippos Paraschos[2], Jongho Park[125,19], Harriet Parsons[24,25], Nimesh Patel[9], Ue-Li Pen[19,32,126,127,128], Dominic W. Pesce[9,8], Vincent Piétu[31], Richard Plambeck[129], Aleksandar PopStefanija[35], Oliver Porth[112,4], Felix M. Pötzl[130,2], Ben Prather[21], Jorge A. Preciado-López[32],

⋆ Corresponding author: anne-kathrin.baczko@chalmers.se
⋆⋆ NASA Hubble Fellowship Program, Einstein Fellow.

Giacomo Principe[131,132,103], Dimitrios Psaltis[124], Hung-Yi Pu[133,134,19], Venkatessh Ramakrishnan[20,135,136], Ramprasad Rao[9], Mark G. Rawlings[137,24,25], Alexander W. Raymond[8,9], Angelo Ricarte[8,9], Bart Ripperda[126,138,127,32], Freek Roelofs[9,8,13], Alan Rogers[14], Cristina Romero-Cañizales[19], Arash Roshanineshat[10], Helge Rottmann[2], Alan L. Roy[2], Ignacio Ruiz[118], Chet Ruszczyk[14], Kazi L. J. Rygl[105], Salvador Sánchez[118], David Sánchez-Argüelles[76,77], Miguel Sánchez-Portal[118], Mahito Sasada[139,85,140], Kaushik Satapathy[10], Tuomas Savolainen[141,136,2], F. Peter Schloerb[35], Jonathan Schonfeld[9], Karl-Friedrich Schuster[31], Lijing Shao[84,2], Zhiqiang Shen[44,45], Des Small[93], Bong Won Sohn[36,37,46], Jason SooHoo[16], León David Sosapanta Salas[112], Kamal Souccar[35], Joshua S. Stanway[143], He Sun[144,145], Fumie Tazaki[85], Alexandra J. Tetarenko[146], Paul Tiede[9,8], Remo P. J. Tilanus[10,13,88,147], Michael Titus[14], Pablo Torne[118,2], Teresa Toscano[16], Efthalia Traianou[16,2], Tyler Trent[10], Sascha Trippe[148], Matthew Turk[48], Ilse van Bemmel[93], Huib Jan van Langevelde[93,88,149], Daniel R. van Rossum[13], Jesse Vos[13], Jan Wagner[2], Derek Ward-Thompson[143], John Wardle[150], Jasmin E. Washington[10], Jonathan Weintroub[8,9], Robert Wharton[2], Kaj Wiik[151], Gunther Witzel[2], Michael F. Wondrak[13,152], George N. Wong[153,42], Qingwen Wu[154], Nitika Yadlapalli[26], Paul Yamaguchi[9], Aristomenis Yfantis[13], Doosoo Yoon[112], André Young[13], Ken Young[9], Ziri Younsi[155,4], Wei Yu[9], Feng Yuan[156], Ye-Fei Yuan[157], J. Anton Zensus[2], Shuo Zhang[158], Guang-Yao Zhao[16,2]

*(Affiliations can be found after the references)*

**ABSTRACT**

*Context.* Many active galaxies harbor powerful relativistic jets, however, the detailed mechanisms of their formation and acceleration remain poorly understood.

*Aims.* To investigate the area of jet acceleration and collimation with the highest available angular resolution, we study the innermost region of the bipolar jet in the nearby low-ionization nuclear emission-line region (LINER) galaxy NGC 1052.

*Methods.* We combined observations of NGC 1052 taken with VLBA, GMVA, and EHT over one week in the spring of 2017. Our study is focused on the size and continuum spectrum of the innermost region containing the central engine and the footpoints of both jets. We employed a synchrotron-self absorption model to fit the continuum radio spectrum and we combined the size measurements from close to the central engine out to ~1 pc to study the jet collimation.

*Results.* For the first time, NGC 1052 was detected with the EHT, providing a size of the central region in-between both jet bases of 43 μas perpendicular to the jet axes, corresponding to just around 250 $R_\mathrm{S}$ (Schwarzschild radii). This size estimate supports previous studies of the jets expansion profile which suggest two breaks of the profile at around $3\times10^3$ $R_\mathrm{S}$ and $1\times10^4$ $R_\mathrm{S}$ distances to the core. Furthermore, we estimated the magnetic field to be 1.25 Gauss at a distance of 22 μas from the central engine by fitting a synchrotron-self absorption spectrum to the innermost emission feature, which shows a spectral turn-over at ~130 GHz. Assuming a purely poloidal magnetic field, this implies an upper limit on the magnetic field strength at the event horizon of $2.6\times10^4$ Gauss, which is consistent with previous measurements.

*Conclusions.* The complex, low-brightness, double-sided jet structure in NGC 1052 makes it a challenge to detect the source at millimeter (mm) wavelengths. However, our first EHT observations have demonstrated that detection is possible up to at least 230 GHz. This study offers a glimpse through the dense surrounding torus and into the innermost central region, where the jets are formed. This has enabled us to finally resolve this region and provide improved constraints on its expansion and magnetic field strength.

**Key words.** methods: observational – techniques: high angular resolution – techniques: interferometric – galaxies: active – galaxies: jets – galaxies: Seyfert

## 1. Introduction

Bipolar extragalactic jets are the most striking features of radio loud active galactic nuclei (AGNs). While propagating up to kiloparsec scales with opening angles smaller than a degree, the collimation and acceleration of the jets takes place within the first parsecs to the central engine. A more accurate description on the launching and collimation region through observations is needed to properly understand the underlying physical processes. This allows us to distinguish between different theoretical models as described, for instance, by Blandford & Znajek (1977) and Blandford & Payne (1982).

To constrain the mechanisms behind collimation at these different scales, the region of acceleration and collimation has to be investigated with the highest achievable resolution through millimeter (mm) Very Long Baseline Interferometry (VLBI). The first studies of the innermost jet region in AGNs with the Event Horizon Telescope (EHT) are very promising. Sub-milliarcsecond jet structures resolved with the EHT were already reported in 3C 279 (Kim et al. 2020), Centaurus A (Janssen et al. 2021), J1924-2914 (Issaoun et al. 2022), NRAO 530 (Jorstad et al. 2023), and 3C 84 (Paraschos et al. 2024). In particular, a bipolar jet base structure of a radio galaxy was revealed in the case of Centaurus A. Only bipolar jets offer the opportunity to also study the symmetric evolution of AGN jets as is assumed within the standard model for AGNs.

Most AGNs with bipolar jets are faint given the small impact of relativistic effects on their brightness. The few cases studied deviate from the overall picture obtained from studies of one-sided blazars, whose jets point towards the observer. Blazar jets typically have a transition from a parabolic collimating jet to a freely expanding conical jet at distances of ~$10^4$–$10^6$ Schwarzschild radii, $R_\mathrm{S}$, (e.g., Kovalev et al. 2020). On the other hand, studies of the expansion profile for strongly misaligned jets do not follow the general trend of Blazar jets. For example NGC 315 shows a break at closer distances of ~$5\times10^3 R_\mathrm{S}$ (Boccardi et al. 2021), or 3C 84 with a nearly cylindrical instead of a parabolic expansion (see, e.g., Giovannini et al. 2018; Nagai et al. 2014).

The nearby (19.23 ± 0.14 Mpc; Tully et al. 2013) low-luminosity AGN (LLAGN) NGC 1052 serves as a bridge between accretion dominated sources (e.g., Sgr A*) and jet-dominated sources (e.g., M 87, 3C84, or Cyg A). It is also often referred to as the prototype low-ionization nuclear emission-line region (LINER) galaxy. X-ray observations suggest that an advection dominated accretion flow (ADAF) is embedded in a truncated accretion disk (AD, see, e.g., Falocco et al. 2020; Reb et al. 2018). It hosts a supermassive black hole (SMBH) of $\simeq 10^{8.2}\,M_{\odot}$ (Woo & Urry 2002). It is one of the few AGNs revealing two jets, one eastward and one westward, which are oriented close to the plane of the sky. A dense torus blocks our view onto the central region at centimeter (cm) wavelengths (e.g., Vermeulen et al. 2003; Kameno et al. 2003; Kadler et al. 2004b; Brenneman et al. 2009). The $H_2O$ maser emission is associated with the torus (Claussen et al. 1998; Kameno et al. 2005; Sawada-Satoh et al. 2008). Observations at millimeter (mm) wavelengths peer through the absorbing structure and reveal a central bright emission feature, which is isolated from the jets at 86 GHz (Baczko et al. 2016a). This first detection of the extended structure at 86 GHz with the Global millimeter VLBI Array (GMVA) allowed for an estimation of the magnetic field at 1 Schwarzschild radius, $R_S$, setting it at a maximum of $10^4$ Gauss. In contrast to most AGNs, the jets in NGC 1052 do not show a parabolic expansion; however, they both evolve with a close-to cylindrical profile, which changes to a close-to conical collimation at $\sim 10^4 R_S$ (Baczko et al. 2022; Nakahara et al. 2020). A recent study suggests strong interaction between jet and torus, which collimates the inner jet and heats the dusty torus (Kameno et al. 2023). A kinematic study at 43 GHz found higher velocities for the eastern (approaching) jet as compared to the western (receding) jet, namely, $\beta_{ej} = 0.529 \pm 0.038$ and $\beta_{wj} = 0.343 \pm 0.037$ in units of the light speed, $c$ (Baczko et al. 2019).

To further investigate the acceleration and collimation zone in NGC 1052 we observed the source at 230 GHz with the EHT and at 86 GHz with the GMVA in 2017. We combined these new observations with multi-frequency VLBA observations (Baczko et al. 2022) close to the EHT campaign in 2017 to study the continuum spectral behavior of the innermost region around the central engine. The GMVA and EHT observations were performed close in time (within one week) to minimize effects from flux density and structural variability.

The paper is organized as follows. In Sect. 2, we present the observations and their data reduction. We describe our methods to obtain size estimates of jet emission features and continuum spectral fitting in Sect. 3. In Sect. 4, we discuss our results in a broader context, giving estimates of the magnetic field close to the central engine, and we compare our size estimates of the jet width with the collimation profile analyzed in Baczko et al. (2022). Finally, Sect. 5 summarizes our findings.

## 2. Observations and data reduction

### 2.1. EHT observation

The EHT observed NGC 1052 at 230 GHz on two consecutive nights on April 6 and April 7, 2017, as a project to further investigate the magnetic field in the innermost region in NGC 1052 (ALMA proposal code 2016.1.01290.V). The source was observed with a part of the EHT array, composed of the Atacama Large Millimeter/submillimeter Array (ALMA, observing as a phased array, Goddi et al. 2019); the IRAM 30 m telescope (PV) in Spain; the *James Clerk Maxwell* Telescope (JCMT) and the Submillimeter Array (SMA) in Hawai'i; and the South Pole Telescope (SPT) in Antarctica. The total time on source was 24 min on Apr 6 (4 scans) and 30 min on Apr 7 (5 scans). Quasar J0132-1654 and blazar J0006-0623 were used as calibrators. The full EHT array setup is detailed in EHTC (2019a).

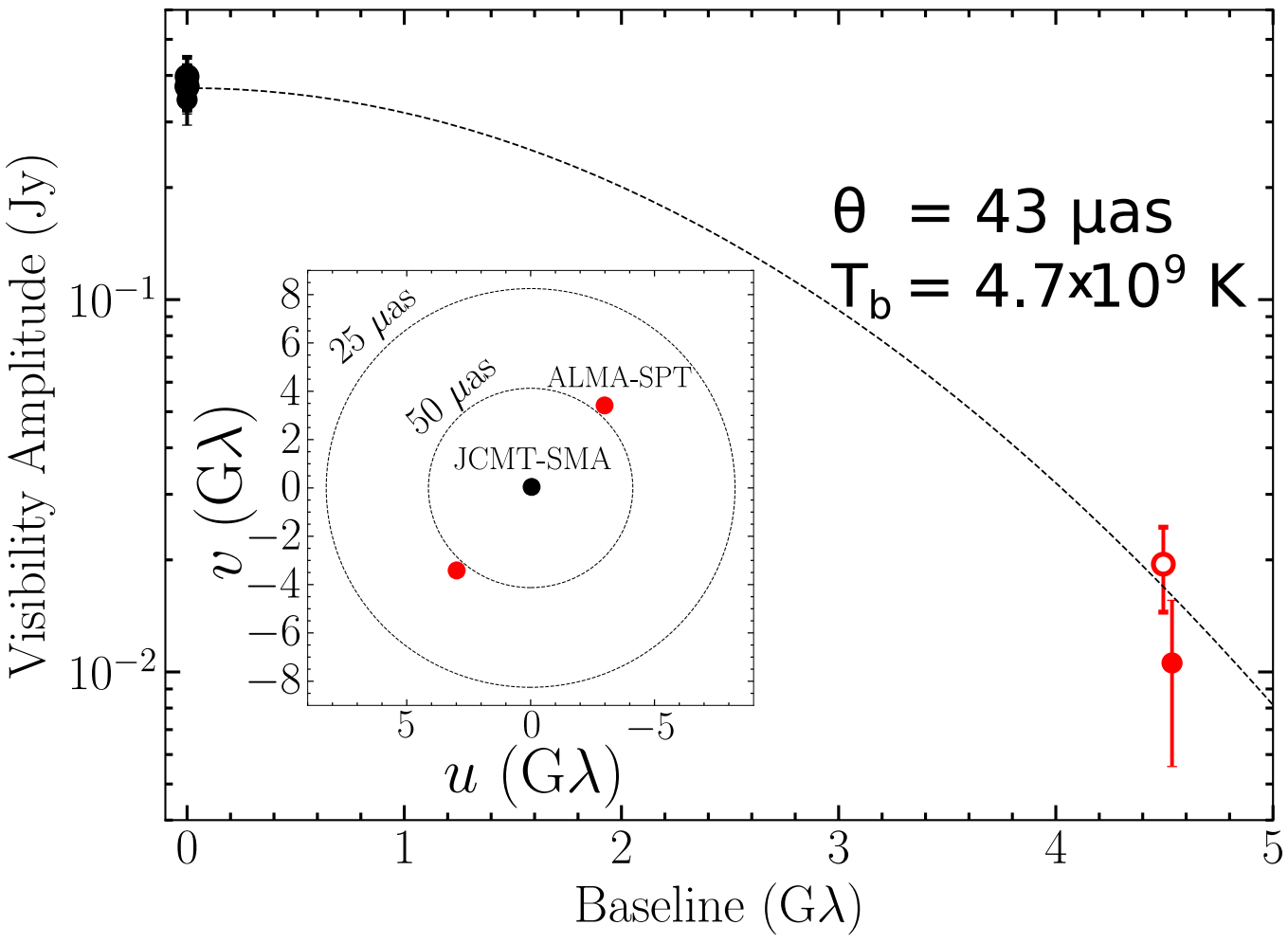

**Fig. 1.** Visibility amplitude and $(u, v)$ coverage (inlet) of EHT observation of NGC 1052 in 2017. The shown brightness temperature is the observed one. The coverage combines detections obtained on April 6 (ALMA-SPT, red markers) and on April 7 (JCMT-SMA, black markers). Empty and filled markers indicate 227.1 and 229.1 GHz bands, respectively.

The raw data were recorded at a rate of 32 Gbps in two ~2 GHz bands centered at 227.1 and 229.1 GHz. Recorded signals were correlated at the MIT Haystack Observatory and the Max-Planck-Institut für Radioastronomie, Bonn. Subsequent data calibration procedures are described in EHTC (2019b, 2022) that relied on a custom-built `EHT-HOPS` fringe-fitting and flux density calibration pipeline (Blackburn et al. 2019), as well as a parallel CASA-based pipeline used for an independent verification (Janssen et al. 2019). Only a very limited number of significant detections on JCMT-SMA (intra-site baseline of $\sim 0.1 M\lambda$ projected length detections at S/N ~ 20) and ALMA-SPT (long baseline of $\sim 4.5 G\lambda$ projected length, detections at S/N ~ 10) was found through incoherent averaging of the visibility amplitudes (Thompson et al. 2017). This is due to particularly low brightness of the source at the time of the EHT observations (0.4 Jy reported by the ALMA-array; Goddi et al. 2021), poor conditions at ALMA observing in the late morning and at a low inclination (Goddi et al. 2019) and poor performance of the individual EHT stations. The poor observing conditions are also reflected in the low quality of calibrators' data, although a larger number of confident detections was obtained given their higher observed flux density. The final visibility amplitude and $(u, v)$ coverage is shown in Fig. 1, providing a total compact flux density of 0.35 ± 0.05 Jy. The visibilities are consistent with a circular Gaussian model with a full width at half maximum (FWHM) of 43 µas, resulting in a brightness temperature estimate of $T_b = (4.7 \pm 0.8) \times 10^9$ K and a circular Gaussian size estimate of (43±6) µas, where uncertainties are dominated by the absolute flux density calibration uncertainties (see EHTC 2019b).

### 2.2. GMVA observation

NGC 1052 was observed with the GMVA at 86 GHz on March 31, 2017 for a total time of 13 hours. The sources 3C 84 and 0224+069 served as fringe finders. The array consisted of 8 VLBA antennas (NL, MK, LA, KP, FD, BR, PT, and OV), Green Bank Telescope (GBT), Yebes, Pico Veleta, Onsala, Metsahovi, and Effelsberg. As is typical in GMVA observations, the scheduling switched between scans on targets and on calibrators, with an average of 7 min on target sources. This allows for a delay and rate transfer from the calibrator to the target source.

The GMVA observation was calibrated using a standard approach in the NRAO Astronomical Imaging Processing System (AIPS, e.g., Baczko et al. 2016b) with a few alterations for fringe fitting. Because high-frequency GMVA observations are very sensitive to the weather, special emphasis was placed on the amplitude calibration, which included a correction for the opacity. The fringe fitting stage revealed unusually noisy data for all sources; specifically for NGC 1052, but also on longer baselines for the calibrators. NGC 1052 itself is relatively faint and has a very complex structure which makes the fringe fitting stage challenging. In addition, there are no bright calibration sources nearby with 0224+069 being the closest one, which is only slightly brighter and more compact compared to NGC 1052. We used GBT as reference antenna for fringe fitting. After finding delays and rates for all sources with the task FRING in AIPS, we interpolated the solutions between sources using the options dobtween=1 and doblank=1 in the AIPS task CLCAL to recover visibilities on more baselines for the target source NGC 1052. This approach can lead to corrupted data remaining after this calibration. Hence, we applied a thorough data flagging in order to remove outliers identified in the radial and baseline visibility plots.

During the imaging using the CLEAN algorithm in DIFMAP (Shepherd et al. 1994), we deviated slightly from the standard procedure. Due to the small amount of finally calibrated data self-flagging was turned off and spurious data, outliers in the Visibility plot were flagged by hand after careful inspection. The final image is shown in Fig. 2 and shows a core dominated morphology with extended jet structure within the inner 1.5 mas. The map is centered on the brightest pixel. The final image parameters are listed in Table 1. The best quality image that recovered the most extended structure and the lowest noise level was obtained using uniform weighting (*uvw 2,-1*) in DIFMAP. Lastly, we modeled the source using circular Gaussian Modelfit components in DIFMAP with six components. In the following, we only use the modelfit of the central component. The exact properties of the modelfit components of the extended structure are uncertain due to the insufficient data quality.

### 2.3. VLBA observations

On April 4, 2017, NGC 1052 was observed at six frequencies from 1.4 GHz to 43 GHz with the VLBA for a total of 11 h. The observations were planned to maximize $(u, v)$-coverage by switching between all 6 frequencies throughout the whole observing run. The properties of the observations and the data reduction procedures are presented in detail in Baczko et al. (2022), while basic map parameters for the 22 and 43 GHz maps used in this publication are listed in Table 1. The data were calibrated using the ParselTongue Python interface (Kettenis et al. 2006) to AIPS to apply exactly the same routines to all frequencies. We used the map from an initial calibration round to enlarge the fringe detection rate during a global fringe. For our study of the innermost region, where both jets are formed, we focus only on the higher-frequency observations by combining the EHT and GMVA results with the 22 and 43 GHz VLBA observations. In the case of very small errors for small component sizes, we set a lower boundary to the positional error equal to the beam size divided by 10, following the approach from Lister et al. (2009).

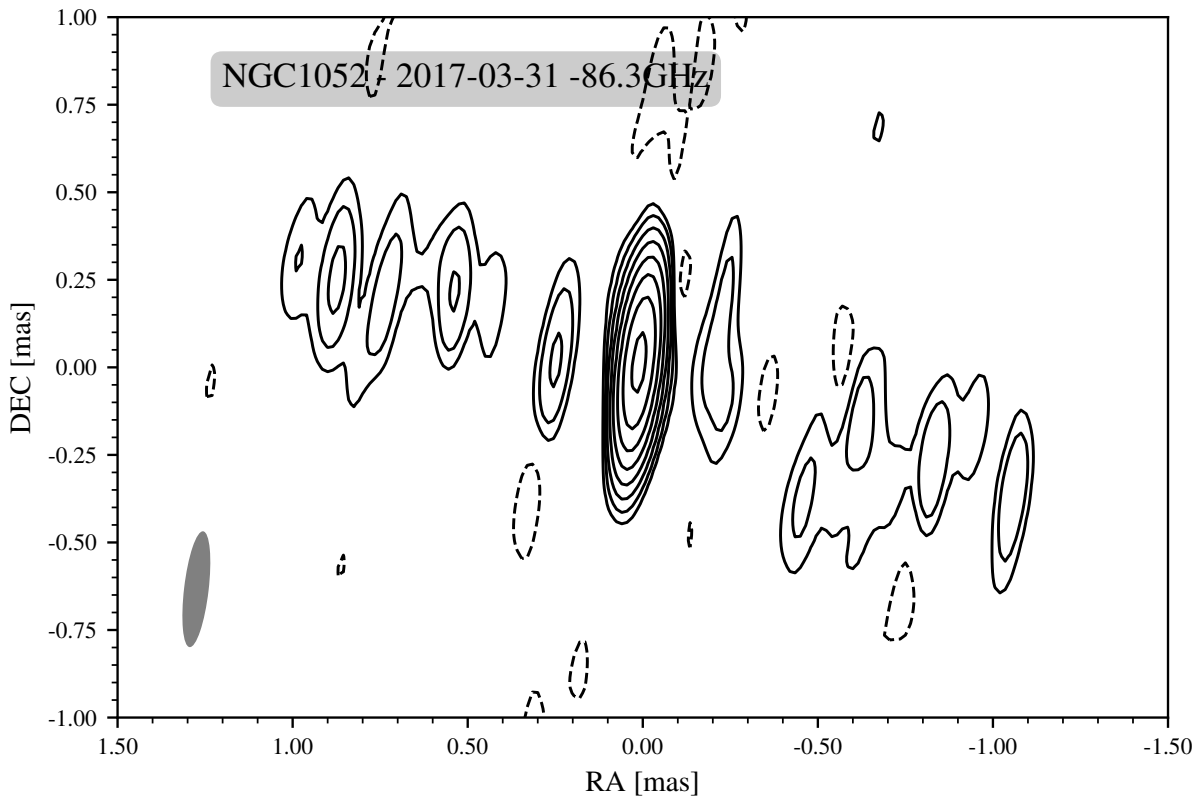

**Fig. 2.** Uniformly weighted CLEAN image of NGC 1052 at 86 GHz, observed with the GMVA on March 31, 2017. This image is centered on the brightest spot of the map. The contours start at two times the noise level of 2.05 mJy/beam, respectively and increase logarithmically by factors of 2. The first negative contours at noise level are shown in dashed lines. The dynamic range of the images is 450. The CLEAN beam is plotted in the lower left corner. The parameters of the final CLEAN image is listed in Table 1.

To study the continuum spectrum from 22 to 230 GHz, we modeled the VLBA maps with 2D-Gaussian functions (model components) for all VLBA maps. Due to the high S/N of the maps, uncertainty estimates for the model components parameters following Fomalont (1999) resulted in unreasonable small uncertainties. As a consequence, we assumed conservative uncertainties on the model component width equal to 1/5th of the beam size and on component position equal to 1/10th of the beam size, following the approach in Lister et al. (2009). Figure 3 shows the model components plotted on top of the clean contour maps for the 22, 43, and 86 GHz. The parameters of the model components for all VLBA observations are available on CDS and on Zenodo[1].

## 3. Results

### 3.1. Aligning images from 1.5 to 86 GHz.

To study the innermost jet-forming region in NGC 1052 we combined the 22 and 43 GHz VLBA observations with the 86 GHz GMVA and 230 GHz EHT observation. The VLBA maps had been aligned using 2D-cross correlation according to Baczko et al. (2022) to shift clean maps and model components of the VLBA observations. While we used the same shift parameters as obtained from Baczko et al. (2022), here we use the model components to verify the plausability of the alignment. As can be seen in Fig. 3 the assumed shifts are also well aligned with the model components, allowing us to trace the same emission regions over frequency. To further support our successful alignment, we also applied fits to the continuum spectrum of the identified components with a simple power-law and SSA spectrum, where appropriate. The successful fitting results support the assumed alignment (supporting figures are provided on Zenodo).

By comparing the 43 and 86 GHz clean images and model fits, we identified the central 86 GHz component with component *A15*, which is the first component of the western jet at

[1] https://zenodo.org/records/13868054

**Table 1.** Image parameters for all analyzed VLBA observations from April 4, 2017 with natural weighting and GMVA observation from March 31, 2017 with uniform weighting.

| Array | $\nu$ [GHz] | RMS [(1a)] $[\frac{\text{mJy}}{\text{beam}}]$ | RMS [(1b)] $[\frac{\text{mJy}}{\text{beam}}]$ | $S_{\text{peak}}$ [(2)] $[\frac{\text{mJy}}{\text{beam}}]$ | $S_{\text{tot}}$ [(3)] [Jy] | $b_{\text{maj}}$ [(4)] [mas] | $b_{\text{min}}$ [(5)] [mas] | PA [(6)] [°] | DR [(7)] |
|---|---|---|---|---|---|---|---|---|---|
| GMVA | 86.2 | 2.05 | 1.4 | 0.63 | 0.81 | 0.33 | 0.07 | −6.7 | 450:1 |
| VLBA | 43.1 | 0.10 | 0.10 | 0.15 | 0.59 | 0.45 | 0.19 | −4.4 | 1500:1 |
| VLBA | 22.2 | 0.08 | 0.09 | 0.18 | 0.80 | 0.86 | 0.34 | −5.0 | 2000:1 |

**Notes.** [(1)](a) Root-mean-square (rms) noise level of image, (b) rms inside a structure-free window far away from the source structure [(2)]Peak brightness, [(3)]Total recovered flux density, [(4)], [(5)], [(6)]Major, minor axes and major axis position angle of the restoring beam, [(7)]Dynamic range: ratio between the map peak and the rms inside a structure-free window far away from the source structure.

43 GHz. This interpretation is in alignment with a multi-year 43 GHz study of NGC 1052, whereby the central, brightest feature in the 43 GHz maps coincides with the kinematic center of the source (Baczko et al. 2019). This is further supported by the previous identification of the central features in quasi-simultaneous GMVA observations at 43 and 86 GHz in 2004 (Baczko et al. 2016a). Based on this assumption, we derived the shift of (0.098 and 0.176) mas in DEC and RA between component *A15* at 43 GHz and 86 GHz. Component *A15* is slightly offset from the map peak at 86 GHz, which is most likely a result of the region being unresolved at 86 GHz. This additional shift was then applied to all VLBA images. The clean maps and components including the shifts from 22 to 86 GHz are shown in Fig. 3. Based on this alignment, we identified the model components between the VLBA observations and the GMVA. The new GMVA image is core-dominated, as is the first 3 mm image of NGC 1052 (Baczko et al. 2016a), with about 80% of the total flux density inside the innermost unresolved feature, corresponding to 664 mJy. In agreement with previous interpretations, we assume this central feature to be located at the central engine. Furthermore, we assume that the effect of free-free absorption (FFA) from the surrounding torus is negligible above 43 GHz. Thus, we conclude that the compact emission observed with the EHT (from visibility analysis only; see Sect. 2) also corresponds to the central engine location, namely, *A15*.

### 3.2. Continuum spectrum of A15

After identifying component *A15* from 43 to 230 GHz, we studied the spectral shape of this component, as illustrated in Fig. 4. It hints towards a spectral turnover between 86 and 230 GHz. To improve the spectral fitting we added another measurement of the flux density at 22 GHz. Indeed, the component at 22 GHz lies in between components *A15* and *B6* at 43 GHz. Assuming this component at 22 GHz is a blend of *A15* and *B6* at 43 GHz we can estimate the flux density of component *A15* to be equal to 50% of this components flux density. This provides us with a fourth point for the spectral fitting of 17.5 mJy for component *A15* at 22 GHz. As many assumptions have gone into this estimate, we have assumed a large uncertainty on this data point of 50%. For 43, 86, and 230 GHz, we adopted a typical, conservative flux density uncertainty of 15%. This is slightly higher as the 10% flux density uncertainty typical for the VLBA , as estimated as most conservative for 15 and 22 GHz VLBA observations of MOJAVE sources (Homan et al. 2002). In a detailed study of 29 epochs of 43 GHz VLBA observations of NGC 1052 we found a typical uncertainty of 14% by means of gscale statistics in difmap (Baczko et al. 2019).

For the fit, we employed a basic synchrotron self-absorption (SSA) spectrum following Eq. (1) (Türler et al. 1999), but leaving the optically thin and optically thick spectral indices as well as the peak frequency and brightness as free parameters, as follows:

$$S_\nu = S_{\text{Pm}} \left(\frac{\nu}{\nu_{\text{m}}}\right)^{\alpha_{\text{thick}}} \frac{1-\exp\left(-\tau_{\text{m}}\left(\frac{\nu}{\nu_{\text{m}}}\right)^{\alpha_{\text{thin}}-\alpha_{\text{thick}}}\right)}{1-\exp(-\tau_{\text{m}})} \tag{1}$$

where $\tau_{\text{m}}$ is the optical depth defined as:

$$\tau_{\text{m}} = \frac{3}{2}\sqrt{1-\left(\frac{8\alpha_{\text{thin}}}{3\alpha_{\text{thick}}}\right)} - 1 \tag{2}$$

This results in a peak frequency at 126 GHz with 1.75 Jy peak brightness, a thin spectral index of $\alpha_{\text{thin}} = -3.7$, and a thick spectral index $\alpha_{\text{thick}} = 3.3$, which exceeds the limit of 2.5 for SSA[2]. Free-free absorption is known to have a significant impact on the detected jet emission at frequencies below 43 GHz in NGC 1052 (Vermeulen et al. 2003; Kameno et al. 2003; Kadler et al. 2004b) and is a likely explanation of the steep optically thick spectral index. If our assumption of the flux density at 22 GHz is wrong and the flux density for *A15* is higher, then the best fit remains about the same, with $\nu_{\text{m}} \simeq 130$ GHz; it is only if the flux density is much lower (around 5 mJy) that spectral peak goes towards lower frequencies $\nu_{\text{m}} \simeq 100$ GHz (further $\alpha_{\text{thick}} \simeq 4$, $\alpha_{\text{thin}} \simeq -1$, and $S_{\text{m}} \simeq 0.7$). Based on the spectral index between 22 and 43 GHz derived from the clean images (compare figures on Zenodo), the flux density of *A15* at 22 GHz is most likely not as low as 5 mJy; rather, it is closer to the assumed 17.5 mJy. This results in an apparent flattening of the spectrum below 43 GHz, which deviates from our simple assumption of the spectral slope.

## 4. Discussion

### 4.1. The spectral shape of A15

The spectral fit to component *A15* suggests a turnover at ≃130 GHz, with a steep, optically thin spectral index of around −3.7. This is larger than the negative spectral index of ∼ − 1 at 220 GHz reported by Goddi et al. (2021), but close to the free-free absorbed synchrotron spectra of the jet with an index of −3 indicated by the Band 3 and Band 4 ALMA continuum spectrum (80–142 GHz) (Kameno et al. 2023). Previous multi-epoch observations suggested that the impact of free-free absorption at frequencies above 43 GHz is small and negligible at 86 GHz (Baczko et al. 2019); hence, the spectral shape between 86 GHz and 230 GHz is unlikely to be affected by strong absorption in the torus. However, it results in a larger uncertainty of the flux density of the innermost component at 22 GHz and in a very steep slope between 22 and 86 GHz, exceeding the limit for synchrotron self-absorption (SSA).

[2] Given the limited number of data points, equal to the number of free parameters in the SSA model, the fit does not provide robust estimates of the uncertainties of the fitted parameters.

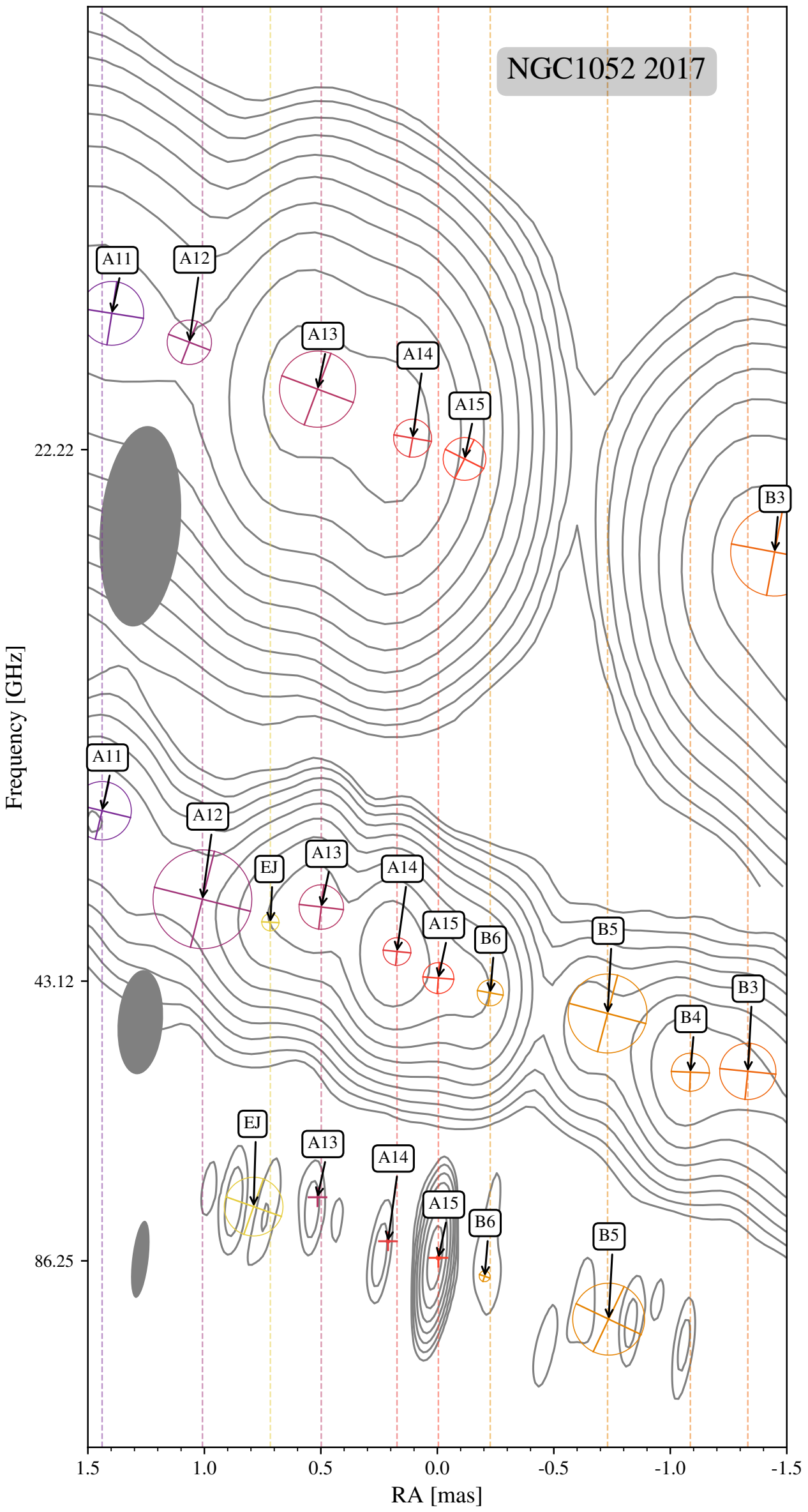

**Fig. 3.** Contour maps of NGC 1052 at 22, 43, and 86 GHz with Gaussian model components plotted on top, the contours start at 3 times the noise level. The DEC scale is equal to the RA scale. The map origin is located at the map peak of the 86 GHz image. VLBA maps were shifted with respect to the 86 GHz image based on the alignment described in Baczko et al. (2022) and by identifying the 86 GHz central component with *A15*. The dashed lines correspond to the component positions at 43 GHz. The components have names assigned as A for the eastern jet and B for the western jet.

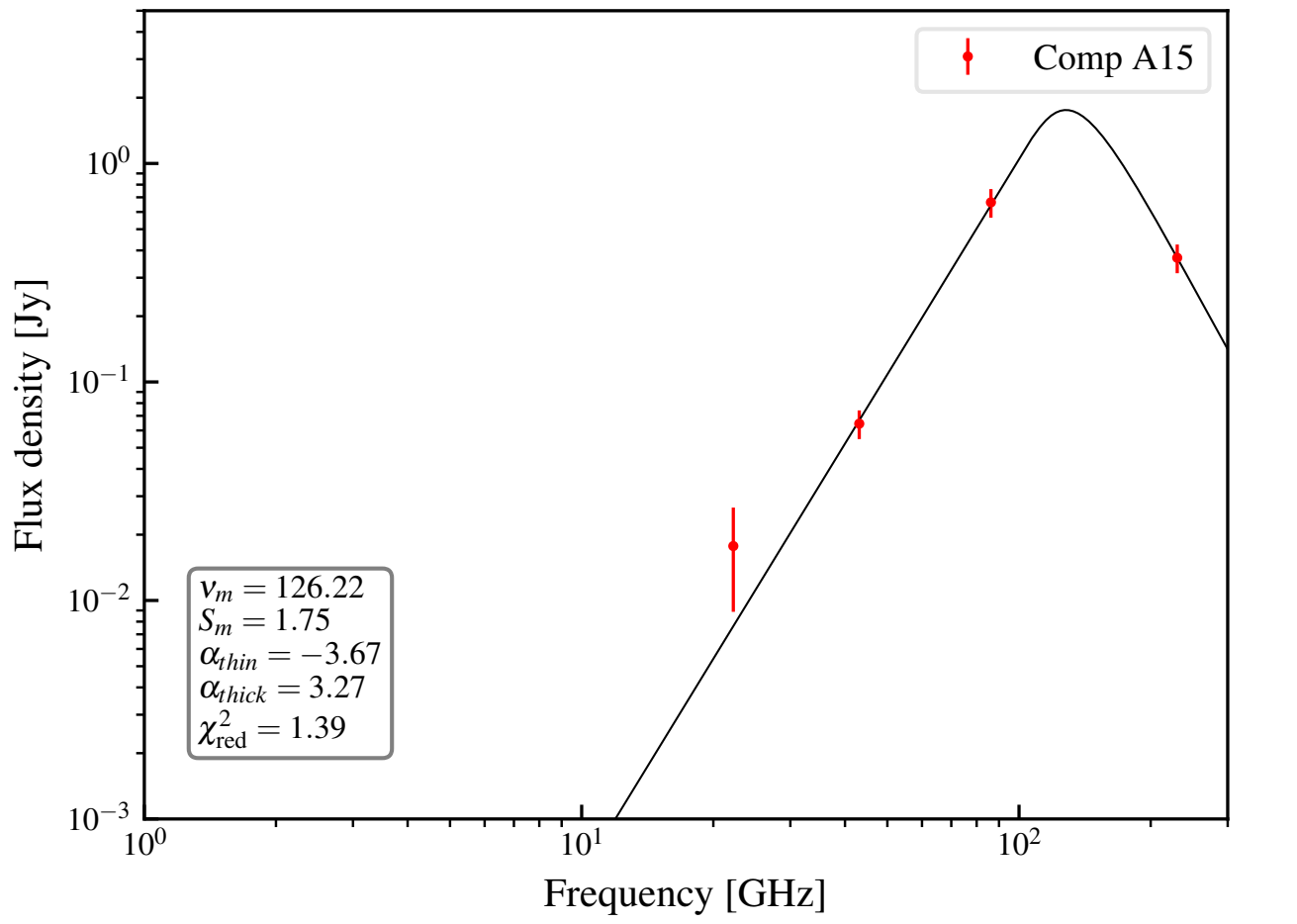

**Fig. 4.** Continuum spectrum of the innermost component *A15* (red dots) and SSA spectral fit with optically thin and thick spectral indices as well as peak frequency and brightness as free parameters (black line). The innermost component A15 has a inverted spectrum below 86 GHz and reveals a turn-over around 130 GHz.

The broad-band spectral energy distribution from $10^8$ Hz up to $10^{15}$ Hz, including high-angular-resolution data, displays a broken power law in the radio-to-UV range with a steep power-law index of −2.6 of the IR-to-UV core continuum (Fernández-Ontiveros et al. 2019). Together with a "harder when brighter" behaviour of the X-ray spectrum (Connolly et al. 2016), suggesting synchrotron self-Compton radiation, and a mild optical extinction – this favours non-thermal emission from a compact jet.

Assuming that the spectral shape around the peak results indeed from SSA, we can estimate the magnetic field at the location of *A15*, assuming equipartition between the magnetic field and the non-thermal electrons (as is shown, e.g., in Baczko et al. 2016a; Kadler et al. 2004a). Assuming a turn-over at 130 GHz, a peak brightness of 1.75 Jy, and a size equivalent to the EHT size of 43 µas, we obtained $T_{\rm B} = 6.9 \times 10^{10}$ K. On this basis, we deduced a magnetic field of ~1.25 G within these inner 43 µas. Given the close distance of the source and the large inclination angle of the jets, we did not correct for the source cosmological redshift and Doppler factor as they are negligible in comparison to the uncertainties of our measurements.

Baczko et al. (2016a) derived a lower limit for the magnetic field based on synchrotron cooling of 6.7 G at a distance of 15 µas (1.5 mpc) from the central engine. Both measurements provide comparable values for the magnetic field strength assuming a toroidal magnetic field between 15 µas and 21.5 µas. Assuming a purely poloidal magnetic field with $B \propto d^{-2}$, with $d$ being the distance to the center, we obtained an upper limit on the magnetic field at the event horizon ( $1\,R_{\rm S}$ distance to the central engine) of $B_{\rm SSA} = 2.6 \times 10^4$ G. Meanwhile, a change from a toroidal B-field distribution ($B \propto d^{-1}$) to a poloidal at a distance of $2\,R_{\rm S}$ to the central engine results in a lower estimate of $B_{\rm SSA} = 392$ G at $1\,R_{\rm S}$. We refer to Baczko et al. (2016b) for a detailed description of the calculations.

The spectral shape might also be explained including non-thermal emission from an ADAF, which has been considered for other LLAGNs. A spectral model combining an ADAF with and without non-thermal electrons and a jet component has been compared to several LLAGNs, such as M 87, M 84, and Cen A, by Bandyopadhyay et al. (2019). This suggests that some source spectra require non-thermal electrons to be considered in the ADAF contribution. Similar studies by Nemmen et al. (2014) had been focused on LINER sources. A visual comparison of the obtained models from these studies with our high-resolution radio data of NGC 1052 suggests that the spectral shape could be explained by a combination of jet and ADAF emission. This is a likely scenario, as previous observations of NGC 1052 already hint towards an ADAF embedded in a truncated accretion disk (Falocco et al. 2020; Reb et al. 2018). Further high-resolution, high-sensitivity data in the mm to sub-mm range are required to better model the broad-band spectrum from the radio to IR and to verify the ADAF contribution. When combined with high-resolution IR to UV data (e.g., Fernández-Ontiveros et al. 2019), a spectral model including ADAF emission ought to be considered to fully describe the core emission.

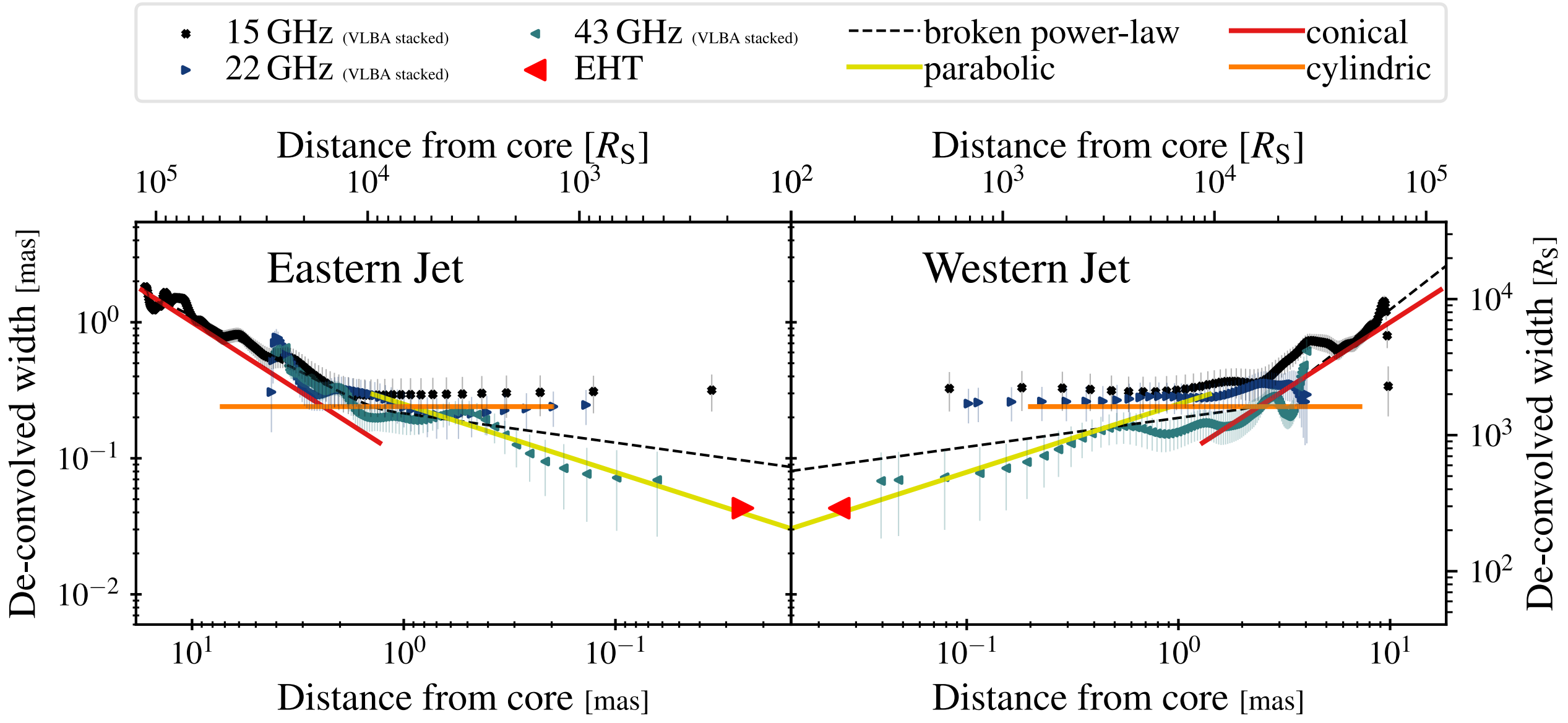

**Fig. 5.** Jet width of stacked 15, 22, and 43 GHz VLBA images reported in Baczko et al. (2022) and 230 GHz (EHT 2017). The black, dotted line denotes a fitted broken power-law to the VLBA images with the power-law indices for the Eastern jet of $k_u = 0.22 \pm 0.06$ and $k_d = 0.80 \pm 0.01$ for upstream and downstream of the break point, respectively, and for the Western jet of $k_u = 0.26 \pm 0.06$ and $k_d = 1.22 \pm 0.05$. Red triangle shows the size of the EHT component at an upper limit on the distance to the center. Orange, yellow, and red lines correspond to (not fitted) power laws with power-law indices of $k = 0$, $k = 0.5$, and $k = 1$, as suggested by the high-frequency data.

The sparse frequency coverage limits the conclusions which can be drawn from the available data sets. Future observations with better sensitivity and further improved uv-coverage at 3 mm and 1.3 mm should make it possible to obtain high-fidelity maps of the source structure at mm wavelengths. The first time detection of NGC 1052 with the EHT at 1.3 mm paves the way for imaging at the highest possible angular (and spatial) resolution. In a future work, we will focus on the combination of our high-resolution VLBI data with previously mentioned archival and possible new millimeter to submillimeter (mm to submm) observations at the highest possible angular resolutions. This will allow us to compare the continuum spectrum with more advanced and complete models, also taking into account contributions from an ADAF.

### 4.2. Unambiguous identification of component A15

Throughout our analysis, we identified the central component at 86 GHz and the EHT detection with component *A15*. This is the most likely identification given the typical symmetry and core dominance of the source structure at 43 GHz, as revealed by multi-year observations (cf. Baczko et al. 2019). However, a second possible identification is with component *B6*. In this scenario, the turn-over around 100 GHz persists; however, the overall spectrum could be fitted successfully with either a power-law or a SSA-spectrum (compare the figures on Zenodo). The spectrum shows a very clear flattening below 43 GHz, as compared to the identification with *A15*, even when considering the large uncertainty of the 22 GHz measurement.

### 4.3. Combined size of the central region

Neither the GMVA observation from 2004 (Baczko et al. 2016a) nor the new observation resolved the innermost feature. The Gaussian modelfit to the GMVA images yields a circular component with a major axis of 20 µas and a brightness temperature of $2.95 \times 10^{11}$ K. This size corresponds to $b_{\rm min}/3.5$ and $b_{\rm maj}/16.5$ for the uniform weighted map. When remaining conservative and assuming a resolution limit of half the beam size, this component is still not resolved in any direction at 86 GHz. However, in accordance with Baczko et al. (2016a), we can assume the resolution limit to give an upper size estimate of the central region along the jet axis as 35 µas. Combining this with the EHT size estimate perpendicular to the jet axis allows us to estimate the size of the central region to $(43 \times 35)$ µas, corresponding to $\sim(280 \times 230)\,R_S$, transversally and parallel to the jet axis.

### 4.4. Collimation profile at the jet base

Based on multi-frequency VLBA observations, the collimation profile in NGC 1052 was found to change from nearly cylindrical to conical expansion at a distance of $10^4\,R_S$ (Baczko et al. 2022). Due to the limited resolution of the highest frequency observation of this study (43 GHz VLBA), it was not possible to infer the jet expansion profile at distances $<10^3\,R_S$. In Fig. 5 we compare our results from the 230 GHz EHT size estimate (red arrows) with the results from Baczko et al. (2022) by plotting the EHT width onto the results obtained from the stacked VLBA images at 15, 22, and 43 GHz. We did not compare with the GMVA modelfit sizes from this study. The innermost jet region is unresolved and would only provide an upper limit, as the outer jet region the dynamic range of our images is worse compared to the measurements from 43 GHz at the same region. For the innermost component *A15,* we assumed an upper limit on the distance corresponding to half the angular EHT resolution of 50 µas. The EHT size measurement supports a scenario where we would expect a second break in the collimation profile at around $3\times10^3\,R_S$, with a steeper power-law index. This was already hinted at when considering the width measurements from the 43 GHz stacked VLBA image in Baczko et al. (2022). Strong heating of the gas around the nucleus in NGC 1052 (as found in ALMA observations) suggests interactions between the jet and the torus (Kameno et al. 2023). In this scenario, the torus could be responsible for a stronger collimation of the jet within the inner $10^4\,R_S$. Furthermore, the different width upwards of the second break as observed at 15 GHz and 22 GHz compared to 43 GHz and 230 GHz suggest a stratified jet in which we observe an outer layer at frequencies of 22 GHz and below. To make these different collimation profiles more visible we add (colored) lines corresponding to power-law indices of $k = 0$ (cylindrical), $k = 0.5$ (parabolic), and $k = 1$ (conical) onto the jet width measurements in Fig. 5, we used the same parameters for the eastern and western jet. The location of the transition point from cylindrical to conical jet collimation profile and the width of the conical region are based on the fitting results from the western jet. The parabolic line is drawn such that it intersects with the fitted broken power law at the point where the 43 GHz width starts to deviate from the fit.

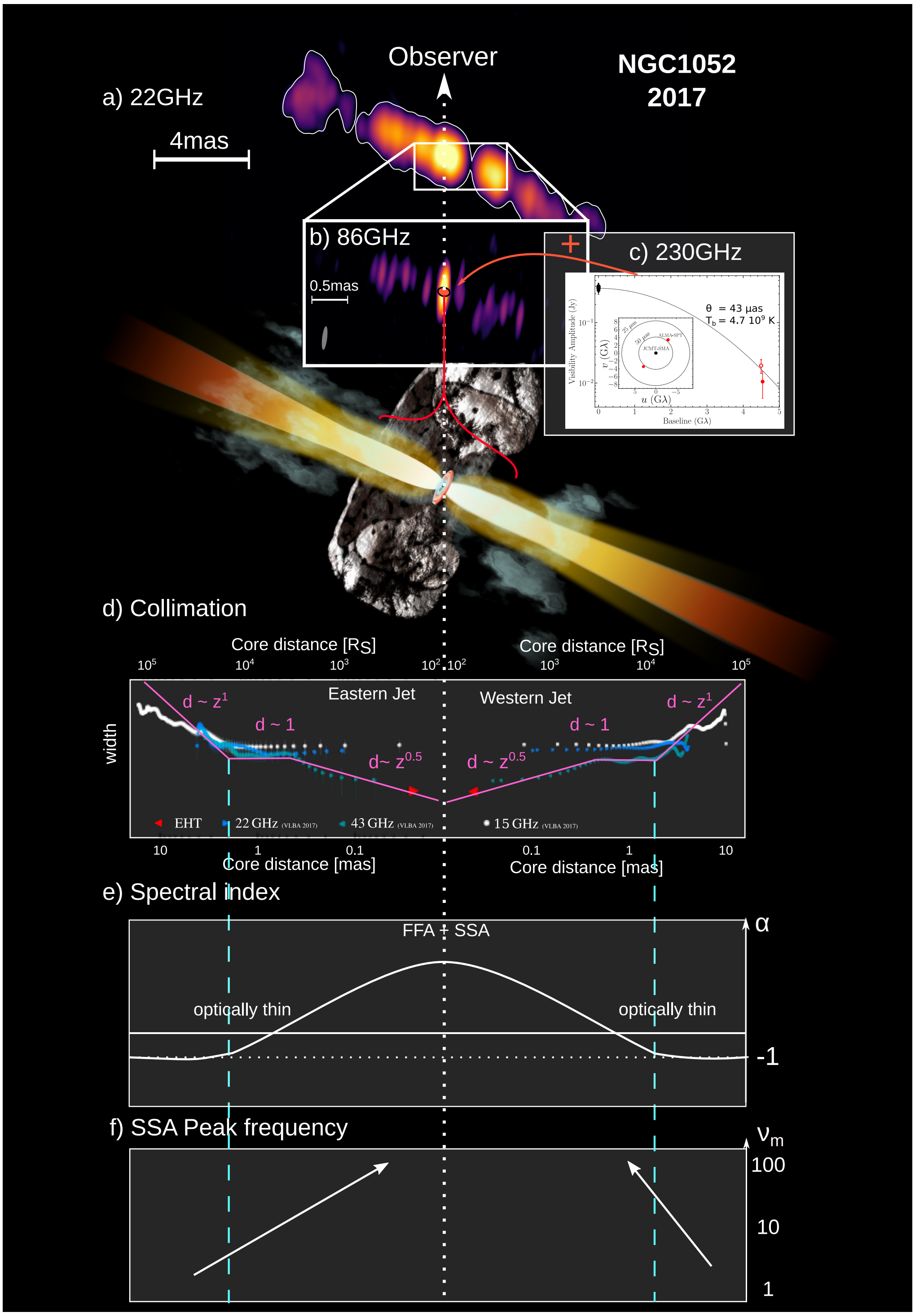

**Fig. 6.** Summary sketch of the inner region in NGC 1052. (a) 2017 22GHz VLBA map with (b) 2017 86 GHz GMVA map zoom in below and (c) EHT visibility plot to the right. Background: Sketch of AGN model, the observer is to the top, the western (right) jet is receding. (d) Jet width at stacked VLBA 22 and 43 GHz, and at 230 GHz and (magenta) power-laws for parabolic, cylindrical, and conical profiles. (e) Sketch visualizing the optically thin and thick regions in the jets. (f) Peak frequency of the SSA-fit moves towards higher frequencies closer to the central engine.

Our observations showed that NGC 1052 can be detected at 230 GHz. Furthermore, the EHT is capable of resolving the central region in NGC 1052 transversally to the jet axes, assuming that the emission detected at 230 GHz corresponds to the innermost component *A15*. The AGN in NGC 1052 is special as it is one of the very few AGNs revealing a double-sided jet at mm wavelengths. The dense molecular surroundings of this source, including the occurrence of water maser emission (Claussen et al. 1998; Kameno et al. 2005; Sawada-Satoh et al. 2008), makes it an extremely interesting target to study in more detail the connection between the host galaxy and the formation and collimation of jets. Stacked multi-frequency VLBA images combined with our results from the EHT suggests a complex transversal structure within the innermost $10^4\,R_\mathrm{S}$. Future full-track EHT observations in combination with higher-resolution GMVA+ALMA observations will have the potential to shed light onto this innermost jet forming region and uncover the true expansion profile. Furthermore, the combination of these high-resolution observations of the area around the central engine with numerical simulation of the same region, also taking into account emission and absorption from both the jets and accretion disk, will allow us to gain a deeper understanding of the formation of AGN jets.

## 5. Summary and conclusions

We present the results from our radio campaign observing NGC 1052 from 1.5 GHz to 230 GHz within a single week time interval. For the first time, it was possible to resolve the innermost central feature in between both jet bases through the 230 GHz EHT observation. Up to 86 GHz, this feature is unresolved. Below and in Fig. 6, we present a summary of our findings.

- For the first time, NGC 1052 was detected at 230 GHz with the EHT on the two baselines ALMA-SPT and JCMT-SMA. From this observation, we infer a size of the emission of the central feature perpendicular to the jet axis of 43 μas, resulting in $T_\mathrm{B} = 4.7 \times 10^9$ K.
- Our new GMVA observation confirms previous results, whereby the central feature constitutes with 0.664 Jy about 80% of the total flux density at 86 GHz.
- Combining the inferred size from the EHT observation and the resolution limit of the GMVA observation provides a size of the central region of (43 × 35) μas, corresponding to ∼(280 × 230) $R_\mathrm{S}$, transverse and parallel to the jet axis.
- By combining these new GMVA and EHT observations with previously published multi-frequency VLBA observations from Baczko et al. (2022), we have traced the innermost emission feature A15 over four frequencies. An SSA-fit to the continuum spectrum of A15 yields a spectral turnover at ∼130 GHz and an upper limit on the magnetic field of 1.25 G. This is consistent with previous measurements and provides an upper limit on the magnetic field at the event horizon (1 $R_\mathrm{S}$ distance to the black hole) of $2.6 \times 10^4$ G, assuming a purely poloidal magnetic field distribution.

Our new observations demonstrate that NGC 1052 can be observed at frequencies up to 230 GHz. We also see that the highest frequencies are required to shed light onto the formation and collimation process of this uncommon double-sided jet system.

## Data availability

The final self-calibrated clean image at 86 GHz and the tables for model fit components are available at the CDS via anonymous ftp to `cdsarc.cds.unistra.fr` (`130.79.128.5`) or via `https://cdsarc.cds.unistra.fr/viz-bin/cat/J/A+A/692/A205`

Appendix B, with additional figures and tables used for this publication, is available on Zenodo at `https://zenodo.org/records/13868054`

*Acknowledgements.* The full acknowledgements are available in Appendix A. The work of MGR is supported by the international Gemini Observatory, a program of NSF NOIRLab, which is managed by the Association of Universities for Research in Astronomy (AURA) under a cooperative agreement with the U.S. National Science Foundation, on behalf of the Gemini partnership of Argentina, Brazil, Canada, Chile, the Republic of Korea, and the United States of America. This work was supported by the National Research Foundation of Korea(NRF) grant funded by the Korea government(MSIT) (RS-2024-00449206). JD is supported by NASA through the NASA Hubble Fellowship grant HST-HF2-51552.001A, awarded by the Space Telescope Science Institute, which is operated by the Association of Universities for Research in Astronomy, Incorporated, under NASA contract NAS5-26555.

[1] Department of Space, Earth and Environment, Chalmers University of Technology, SE-41296 Gothenburg, Sweden
[2] Max-Planck-Institut für Radioastronomie, Auf dem Hügel 69, D-53121 Bonn, Germany
[3] Institut für Theoretische Physik und Astrophysik, Universität Würzburg, Emil-Fischer-Str. 31, D-97074 Würzburg, Germany
[4] Institut für Theoretische Physik, Goethe-Universität Frankfurt, Max-von-Laue-Straße 1, D-60438 Frankfurt am Main, Germany
[5] Departament d'Astronomia i Astrofísica, Universitat de València, C. Dr. Moliner 50, E-46100 Burjassot, València, Spain
[6] Observatori Astronòmic, Universitat de València, C. Catedrático José Beltrán 2, E-46980 Paterna València, Spain
[7] Yale Center for Astronomy & Astrophysics, Yale University, 52 Hillhouse Avenue, New Haven, CT 06511, USA
[8] Black Hole Initiative at Harvard University, 20 Garden Street, Cambridge, MA 02138, USA
[9] Center for Astrophysics | Harvard & Smithsonian, 60 Garden Street, Cambridge, MA 02138, USA
[10] Steward Observatory and Department of Astronomy, University of Arizona, 933 N. Cherry Ave., Tucson, AZ 85721, USA
[11] Data Science Institute, University of Arizona, 1230 N. Cherry Ave., Tucson, AZ 85721, USA
[12] Program in Applied Mathematics, University of Arizona, 617 N. Santa Rita, Tucson, AZ 85721, USA
[13] Department of Astrophysics, Institute for Mathematics, Astrophysics and Particle Physics (IMAPP), Radboud University, PO Box 9010, 6500 GL Nijmegen, The Netherlands
[14] Massachusetts Institute of Technology Haystack Observatory, 99 Millstone Road, Westford, MA 01886, USA
[15] National Astronomical Observatory of Japan, 2-21-1 Osawa, Mitaka, Tokyo 181-8588, Japan
[16] Instituto de Astrofísica de Andalucía-CSIC, Glorieta de la Astronomía s/n, E-18008 Granada, Spain
[17] Department of Physics, Faculty of Science, Universiti Malaya, 50603 Kuala Lumpur, Malaysia
[18] Department of Physics & Astronomy, The University of Texas at San Antonio, One UTSA Circle, San Antonio, TX 78249, USA
[19] Institute of Astronomy and Astrophysics, Academia Sinica, 11F of Astronomy-Mathematics Building, AS/NTU No. 1, Sec. 4, Roosevelt Rd., Taipei 10617, Taiwan, ROC
[20] Astronomy Department, Universidad de Concepción, Casilla 160-C Concepción, Chile
[21] Department of Physics, University of Illinois, 1110 West Green Street, Urbana, IL 61801, USA
[22] Fermi National Accelerator Laboratory, MS209, PO Box 500, Batavia, IL 60510, USA
[23] Department of Astronomy and Astrophysics, University of Chicago, 5640 South Ellis Avenue, Chicago, IL 60637, USA
[24] East Asian Observatory, 660 N. A'ohoku Place, Hilo, HI 96720, USA
[25] James Clerk Maxwell Telescope (JCMT), 660 N. A'ohoku Place, Hilo, HI 96720, USA
[26] California Institute of Technology, 1200 East California Boulevard, Pasadena, CA 91125, USA
[27] Institute of Astronomy and Astrophysics, Academia Sinica, 645 N. A'ohoku Place, Hilo, HI 96720, USA
[28] Department of Physics and Astronomy, University of Hawaii at Manoa, 2505 Correa Road, Honolulu, HI 96822, USA
[29] Department of Physics, McGill University, 3600 rue University, Montréal, QC H3A 2T8, Canada
[30] Trottier Space Institute at McGill, 3550 rue University, Montréal, QC H3A 2A7, Canada
[31] Institut de Radioastronomie Millimétrique (IRAM), 300 rue de la Piscine, F-38406 Saint Martin d'Hères, France
[32] Perimeter Institute for Theoretical Physics, 31 Caroline Street North, Waterloo, ON N2L 2Y5, Canada
[33] Department of Physics and Astronomy, University of Waterloo, 200 University Avenue West, Waterloo, ON N2L 3G1, Canada
[34] Waterloo Centre for Astrophysics, University of Waterloo, Waterloo, ON N2L 3G1, Canada
[35] Department of Astronomy, University of Massachusetts, Amherst, MA 01003, USA
[36] Korea Astronomy and Space Science Institute, Daedeok-daero 776, Yuseong-gu, Daejeon 34055, Republic of Korea
[37] University of Science and Technology, Gajeong-ro 217, Yuseong-gu, Daejeon 34113, Republic of Korea
[38] Kavli Institute for Cosmological Physics, University of Chicago, 5640 South Ellis Avenue, Chicago, IL 60637, USA
[39] Department of Physics, University of Chicago, 5720 South Ellis Avenue, Chicago, IL 60637, USA
[40] Enrico Fermi Institute, University of Chicago, 5640 South Ellis Avenue, Chicago, IL 60637, USA
[41] Department of Space, Earth and Environment, Chalmers University of Technology, Onsala Space Observatory, SE-43992 Onsala, Sweden
[42] Princeton Gravity Initiative, Jadwin Hall, Princeton University, Princeton, NJ 08544, USA
[43] Cornell Center for Astrophysics and Planetary Science, Cornell University, Ithaca, NY 14853, USA
[44] Shanghai Astronomical Observatory, Chinese Academy of Sciences, 80 Nandan Road, Shanghai 200030, PR China
[45] Key Laboratory of Radio Astronomy and Technology, Chinese Academy of Sciences, A20 Datun Road, Chaoyang District, Beijing 100101, PR China
[46] Department of Astronomy, Yonsei University, Yonsei-ro 50, Seodaemun-gu 03722, Seoul, Republic of Korea
[47] Physics Department, Fairfield University, 1073 North Benson Road, Fairfield, CT 06824, USA
[48] Department of Astronomy, University of Illinois at Urbana-Champaign, 1002 West Green Street, Urbana, IL 61801, USA
[49] Instituto de Astronomía, Universidad Nacional Autónoma de México (UNAM), Apdo Postal 70-264, Ciudad de México, Mexico
[50] Research Center for Astronomical Computing, Zhejiang Laboratory, Hangzhou 311100, PR China
[51] Tsung-Dao Lee Institute, Shanghai Jiao Tong University, Shengrong Road 520, Shanghai 201210, PR China
[52] Department of Astronomy and Columbia Astrophysics Laboratory, Columbia University, 500 W. 120th Street, New York, NY 10027, USA
[53] Center for Computational Astrophysics, Flatiron Institute, 162 Fifth Avenue, New York, NY 10010, USA
[54] Dipartimento di Fisica "E. Pancini", Università di Napoli "Federico II", Compl. Univ. di Monte S. Angelo, Edificio G, Via Cinthia, I-80126 Napoli, Italy
[55] INFN Sez. di Napoli, Compl. Univ. di Monte S. Angelo, Edificio G, Via Cinthia, I-80126 Napoli, Italy
[56] Wits Centre for Astrophysics, University of the Witwatersrand, 1 Jan Smuts Avenue, Braamfontein, Johannesburg 2050, South Africa

[57] Department of Physics, University of Pretoria, Hatfield, Pretoria 0028, South Africa
[58] Centre for Radio Astronomy Techniques and Technologies, Department of Physics and Electronics, Rhodes University, Makhanda 6140, South Africa
[59] ASTRON, Oude Hoogeveensedijk 4, 7991 PD Dwingeloo, The Netherlands
[60] LESIA, Observatoire de Paris, Université PSL, CNRS, Sorbonne Université, Université de Paris, 5 place Jules Janssen, F-92195 Meudon, France
[61] JILA and Department of Astrophysical and Planetary Sciences, University of Colorado, Boulder, CO 80309, USA
[62] National Astronomical Observatories, Chinese Academy of Sciences, 20A Datun Road, Chaoyang District, Beijing 100101, PR China
[63] Las Cumbres Observatory, 6740 Cortona Drive, Suite 102, Goleta, CA 93117-5575, USA
[64] Department of Physics, University of California, Santa Barbara, CA 93106-9530, USA
[65] National Radio Astronomy Observatory, 520 Edgemont Road, Charlottesville, USA
[66] Department of Electrical Engineering and Computer Science, Massachusetts Institute of Technology, 32-D476, 77 Massachusetts Ave., Cambridge, MA 02142, USA
[67] Google Research, 355 Main St., Cambridge, MA 02142, USA
[68] Department of History of Science, Harvard University, Cambridge, MA 02138, USA
[69] Department of Physics, Harvard University, Cambridge, MA 02138, USA
[70] NCSA, University of Illinois, 1205 W. Clark St., Urbana, IL 61801, USA
[71] Instituto de Astronomia, Geofísica e Ciências Atmosféricas, Universidade de São Paulo, R. do Matão, 1226, São Paulo, SP 05508-090, Brazil
[72] Dipartimento di Fisica, Universitá degli Studi di Cagliari, SP Monserrato-Sestu km 0.7, I-09042 Monserrato, Italy
[73] INAF – Osservatorio Astronomico di Cagliari, Via della Scienza 5, I-09047 Selargius (CA), Italy
[74] INFN, sezione di Cagliari, I-09042 Monserrato, (CA), Italy
[75] CP3-Origins, University of Southern Denmark, Campusvej 55, DK-5230 Odense M, Denmark
[76] Instituto Nacional de Astrofísica, Óptica y Electrónica, Apartado Postal 51 y 216, 72000 Puebla Pue., Mexico
[77] Consejo Nacional de Humanidades, Ciencia y Tecnología, Av. Insurgentes Sur 1582, 03940 Ciudad de México, Mexico
[78] Key Laboratory for Research in Galaxies and Cosmology, Chinese Academy of Sciences, Shanghai 200030, PR China
[79] Graduate School of Science, Nagoya City University, Yamanohata 1, Mizuho-cho, Mizuho-ku, Nagoya 467-8501, Aichi, Japan
[80] Mizusawa VLBI Observatory, National Astronomical Observatory of Japan, 2-12 Hoshigaoka-cho, Mizusawa, Oshu 023-0861, Iwate, Japan
[81] Trottier Space Institute at McGill, 3550 rue University, Montréal, QC H3A 2A7, Canada
[82] NOVA Sub-mm Instrumentation Group, Kapteyn Astronomical Institute, University of Groningen, Landleven 12, 9747 AD Groningen, The Netherlands
[83] Department of Astronomy, School of Physics, Peking University, Beijing 100871, PR China
[84] Kavli Institute for Astronomy and Astrophysics, Peking University, Beijing 100871, PR China
[85] Mizusawa VLBI Observatory, National Astronomical Observatory of Japan, 2-12 Hoshigaoka, Mizusawa, Oshu, Iwate 023-0861, Japan
[86] Department of Astronomical Science, The Graduate University for Advanced Studies (SOKENDAI), 2-21-1 Osawa, Mitaka, Tokyo 181-8588, Japan
[87] Department of Astronomy, Graduate School of Science, The University of Tokyo, 7-3-1 Hongo, Bunkyo-ku, Tokyo 113-0033, Japan
[88] Leiden Observatory, Leiden University, Postbus 2300, 9513 RA Leiden, The Netherlands
[89] ASTRAVEO LLC, PO Box 1668, Gloucester, MA 01931, USA
[90] Applied Materials Inc., 35 Dory Road, Gloucester, MA 01930, USA
[91] Institute for Astrophysical Research, Boston University, 725 Commonwealth Ave., Boston, MA 02215, USA
[92] Institute for Cosmic Ray Research, The University of Tokyo, 5-1-5 Kashiwanoha, Kashiwa, Chiba 277-8582, Japan
[93] Joint Institute for VLBI ERIC (JIVE), Oude Hoogeveensedijk 4, 7991 PD Dwingeloo, The Netherlands
[94] Department of Physics, Ulsan National Institute of Science and Technology (UNIST), 50 UNIST-gil, Eonyang-eup, Ulju-gun, Ulsan 44919, Republic of Korea
[95] Department of Physics, Korea Advanced Institute of Science and Technology (KAIST), 291 Daehak-ro, Yuseong-gu, Daejeon 34141, Republic of Korea
[96] Kogakuin University of Technology & Engineering, Academic Support Center, 2665-1 Nakano, Hachioji, Tokyo 192-0015, Japan
[97] Graduate School of Science and Technology, Niigata University, 8050 Ikarashi 2-no-cho, Nishi-ku, Niigata 950-2181, Japan
[98] Physics Department, National Sun Yat-Sen University, No. 70, Lien-Hai Road, Kaosiung City 80424, Taiwan, ROC
[99] National Optical Astronomy Observatory, 950 N. Cherry Ave., Tucson, AZ 85719, USA
[100] Department of Physics, The Chinese University of Hong Kong, Shatin, N. T., Hong Kong
[101] School of Astronomy and Space Science, Nanjing University, Nanjing 210023, PR China
[102] Key Laboratory of Modern Astronomy and Astrophysics, Nanjing University, Nanjing 210023, PR China
[103] INAF-Istituto di Radioastronomia, Via P. Gobetti 101, I-40129 Bologna, Italy
[104] Instituto de Física, Pontificia Universidad Católica de Valparaíso, Casilla 4059, Valparaíso, Chile
[105] INAF-Istituto di Radioastronomia & Italian ALMA Regional Centre, Via P. Gobetti 101, I-40129 Bologna, Italy
[106] Department of Physics, National Taiwan University, No. 1, Sec. 4, Roosevelt Rd., Taipei 10617, Taiwan, ROC
[107] Instituto de Radioastronomía y Astrofísica, Universidad Nacional Autónoma de México, Morelia 58089, Mexico
[108] Key Laboratory of Radio Astronomy, Chinese Academy of Sciences, Nanjing 210008, PR China
[109] Yunnan Observatories, Chinese Academy of Sciences, 650011 Kunming, Yunnan Province, PR China
[110] Center for Astronomical Mega-Science, Chinese Academy of Sciences, 20A Datun Road, Chaoyang District, Beijing 100012, PR China
[111] Key Laboratory for the Structure and Evolution of Celestial Objects, Chinese Academy of Sciences, 650011 Kunming, PR China
[112] Anton Pannekoek Institute for Astronomy, University of Amsterdam, Science Park 904, 1098 XH Amsterdam, The Netherlands
[113] Gravitation and Astroparticle Physics Amsterdam (GRAPPA) Institute, University of Amsterdam, Science Park 904, 1098 XH Amsterdam, The Netherlands
[114] Department of Astrophysical Sciences, Peyton Hall, Princeton University, Princeton, NJ 08544, USA
[115] Science Support Office, Directorate of Science, European Space Research and Technology Centre (ESA/ESTEC), Keplerlaan 1, 2201 AZ Noordwijk, The Netherlands
[116] Department of Astronomy and Astrophysics, University of Chicago, 5640 South Ellis Avenue, Chicago, IL 60637, USA
[117] School of Physics and Astronomy, Shanghai Jiao Tong University, 800 Dongchuan Road, Shanghai, 200240, PR China
[118] Institut de Radioastronomie Millimétrique (IRAM), Avenida Divina Pastora 7, Local 20, E-18012 Granada, Spain
[119] National Institute of Technology, Hachinohe College, 16-1 Uwanotai, Tamonoki, Hachinohe City, Aomori 039-1192, Japan

[120] Research Center for Astronomy, Academy of Athens, Soranou Efessiou 4, 115 27 Athens, Greece
[121] Department of Physics, Villanova University, 800 Lancaster Avenue, Villanova, PA 19085, USA
[122] Physics Department, Washington University, CB 1105, St. Louis, MO 63130, USA
[123] Departamento de Matemática da Universidade de Aveiro and Centre for Research and Development in Mathematics and Applications (CIDMA), Campus de Santiago, 3810-193 Aveiro, Portugal
[124] School of Physics, Georgia Institute of Technology, 837 State St NW, Atlanta, GA 30332, USA
[125] School of Space Research, Kyung Hee University, 1732, Deogyeong-daero, Giheung-gu, Yongin-si, Gyeonggi-do 17104, Republic of Korea
[126] Canadian Institute for Theoretical Astrophysics, University of Toronto, 60 St. George Street, Toronto, ON M5S 3H8, Canada
[127] Dunlap Institute for Astronomy and Astrophysics, University of Toronto, 50 St. George Street, Toronto, ON M5S 3H4, Canada
[128] Canadian Institute for Advanced Research, 180 Dundas St West, Toronto, ON M5G 1Z8, Canada
[129] Radio Astronomy Laboratory, University of California, Berkeley, CA 94720, USA
[130] Institute of Astrophysics, Foundation for Research and Technology – Hellas, Voutes 7110, Heraklion, Greece
[131] Dipartimento di Fisica, Università di Trieste, I-34127 Trieste, Italy
[132] INFN Sez. di Trieste, I-34127 Trieste, Italy
[133] Department of Physics, National Taiwan Normal University, No. 88, Sec. 4, Tingzhou Rd., Taipei 116, Taiwan, ROC
[134] Center of Astronomy and Gravitation, National Taiwan Normal University, No. 88, Sec. 4, Tingzhou Road, Taipei 116, Taiwan, ROC
[135] Finnish Centre for Astronomy with ESO, FI-20014 University of Turku, Finland
[136] Aalto University Metsähovi Radio Observatory, Metsähovintie 114, FI-02540 Kylmälä, Finland
[137] Gemini Observatory/NSF's NOIRLab, 670 N. A'ohōkū Place, Hilo, HI 96720, USA
[138] Department of Physics, University of Toronto, 60 St. George Street, Toronto, ON M5S 1A7, Canada
[139] Department of Physics, Tokyo Institute of Technology, 2-12-1 Ookayama, Meguro-ku, Tokyo 152-8551, Japan
[140] Hiroshima Astrophysical Science Center, Hiroshima University, 1-3-1 Kagamiyama, Higashi-Hiroshima, Hiroshima 739-8526, Japan
[141] Aalto University Department of Electronics and Nanoengineering, PL 15500, FI-00076 Aalto, Finland
[142] Physics & Astronomy Department, Rice University, Houston, Texas 77005-1827, USA
[143] Jeremiah Horrocks Institute, University of Central Lancashire, Preston PR1 2HE, UK
[144] National Biomedical Imaging Center, Peking University, Beijing 100871, PR China
[145] College of Future Technology, Peking University, Beijing 100871, PR China
[146] Department of Physics and Astronomy, University of Lethbridge, Lethbridge, Alberta T1K 3M4, Canada
[147] Netherlands Organisation for Scientific Research (NWO), Postbus 93138, 2509 AC Den Haag, The Netherlands
[148] Department of Physics and Astronomy, Seoul National University, Gwanak-gu, Seoul 08826, Republic of Korea
[149] University of New Mexico, Department of Physics and Astronomy, Albuquerque, NM 87131, USA
[150] Physics Department, Brandeis University, 415 South Street, Waltham, MA 02453, USA
[151] Tuorla Observatory, Department of Physics and Astronomy, University of Turku, FI-20014 TURUN YLIOPISTO, Finland
[152] Radboud Excellence Fellow of Radboud University, Nijmegen, The Netherlands
[153] School of Natural Sciences, Institute for Advanced Study, 1 Einstein Drive, Princeton, NJ 08540, USA
[154] School of Physics, Huazhong University of Science and Technology, Wuhan, Hubei 430074, PR China
[155] Mullard Space Science Laboratory, University College London, Holmbury St. Mary, Dorking, Surrey RH5 6NT, UK
[156] Center for Astronomy and Astrophysics and Department of Physics, Fudan University, Shanghai 200438, PR China
[157] Astronomy Department, University of Science and Technology of China, Hefei 230026, PR China
[158] Department of Physics and Astronomy, Michigan State University, 567 Wilson Rd, East Lansing, MI 48824, USA

## Appendix A: Acknowledgements

This research has made use of data obtained with the Global Millimeter VLBI Array (GMVA), which consists of telescopes operated by the MPIfR, IRAM, Onsala, Metsähovi, Yebes, the Korean VLBI Network, the Greenland Telescope, the Green Bank Observatory (GBT) and the Very Long Baseline Array (VLBA). The VLBA and the GBT are facilities of the National Science Foundation operated under cooperative agreement by Associated Universities, Inc. This work made use of the Swinburne University of Technology software correlator, developed as part of the Australian Major National Research Facilities Programme and operated under licence Deller et al. (2011). The data were correlated at the VLBI correlator of the Max-Planck-Institut für Radioastronomie (MPIfR) in Bonn, Germany. This research has made use of eht-imager (Chael et al. 2018) and ParselTongue (Kettenis et al. 2006). This research is supported by the European Research Council advanced grant "M2FINDERS - Mapping Magnetic Fields with INterferometry Down to Event hoRizon Scales" (Grant No. 101018682) and by the Deutsche Forschungsgemeinschaft (DFG, German Research Foundation) - grant 443220636, (FOR5195: Relativistic Jets in Active Galaxies). MP acknowledges support by the Spanish Ministry of Science through Grants PID2019-105510GB-C31/AEI/10.13039/501100011033 and PID2022-136828NB-C43, from the Generalitat Valenciana through grant CIPROM/2022/49, and from the Astrophysics and High Energy Physics programme supported by MCIN with funding from European Union NextGenerationEU (PRTR-C17.I1) and by Generalitat Valenciana through grant ASFAE/2022/005.

The Event Horizon Telescope Collaboration thanks the following organizations and programs: the Academia Sinica; the Academy of Finland (projects 274477, 284495, 312496, 315721); the Agencia Nacional de Investigación y Desarrollo (ANID), Chile via NCN19_058 (TITANs), Fondecyt 1221421 and BASAL FB210003; the Alexander von Humboldt Stiftung; an Alfred P. Sloan Research Fellowship; Allegro, the European ALMA Regional Centre node in the Netherlands, the NL astronomy research network NOVA and the astronomy institutes of the University of Amsterdam, Leiden University, and Radboud University; the ALMA North America Development Fund; the Astrophysics and High Energy Physics programme by MCIN (with funding from European Union NextGenerationEU, PRTR-C17I1); the Black Hole Initiative, which is funded by grants from the John Templeton Foundation (60477, 61497, 62286) and the Gordon and Betty Moore Foundation (Grant GBMF-8273) - although the opinions expressed in this work are those of the author and do not necessarily reflect the views of these Foundations ; the Brinson Foundation; "la Caixa" Foundation (ID 100010434) through fellowship codes LCF/BQ/DI22/11940027 and LCF/BQ/DI22/11940030; Chandra DD7-18089X and TM6-17006X; the China Scholarship Council; the China Postdoctoral Science Foundation fellowships (2020M671266, 2022M712084); Consejo Nacional de Humanidades, Ciencia y Tecnología (CONAHCYT, Mexico, projects U0004-246083, U0004-259839, F0003-272050, M0037-279006, F0003-281692, 104497, 275201, 263356,CBF2023-2024-1102,257435); the Colfuturo Scholarship; the Consejería de Economía, Conocimiento, Empresas y Universidad of the Junta de Andalucía (grant P18-FR-1769), the Consejo Superior de Investigaciones Científicas (grant 2019AEP112); the Delaney Family via the Delaney Family John A. Wheeler Chair at Perimeter Institute; Dirección General de Asuntos del Personal Académico-Universidad Nacional Autónoma de México (DGAPA-UNAM, projects IN112820 and IN108324); the Dutch Research Council (NWO) for the VICI award (grant 639.043.513), the grant OCENW.KLEIN.113, and the Dutch Black Hole Consortium (with project No. NWA 1292.19.202) of the research programme the National Science Agenda; the Dutch National Supercomputers, Cartesius and Snellius (NWO grant 2021.013); the EACOA Fellowship awarded by the East Asia Core Observatories Association, which consists of the Academia Sinica Institute of Astronomy and Astrophysics, the National Astronomical Observatory of Japan, Center for Astronomical Mega-Science, Chinese Academy of Sciences, and the Korea Astronomy and Space Science Institute; The European Research Council (ERC) Synergy Grant "BlackHoleCam: Imaging the Event Horizon of Black Holes" (grant 610058) and Synergy Grant "BlackHolistic: Colour Movies of Black Holes; Understanding Black Hole Astrophysics from the Event Horizon to Galactic Scales" (grant 10107164); the European Union Horizon 2020 research and innovation programme under grant agreements RadioNet (No. 730562), M2FINDERS (No. 101018682) and FunFiCO (No. 777740); the Horizon ERC Grants 2021 programme under grant agreement No. 101040021; the European Research Council for advanced grant 'JETSET: Launching, propagation and emission of relativistic jets from binary mergers and across mass scales' (grant No. 884631); the European Horizon Europe staff exchange (SE) programme HORIZON-MSCA-2021-SE-01 grant NewFunFiCO (No. 10108625); the FAPESP (Fundação de Amparo á Pesquisa do Estado de São Paulo) under grant 2021/01183-8; the Fondo CAS-ANID folio CAS220010; the Generalitat Valenciana (grants APOSTD/2018/177 and ASFAE/2022/018) and GenT Program (project CIDEGENT/2018/021); the Gordon and Betty Moore Foundation (GBMF-3561, GBMF-5278, GBMF-10423); the Institute for Advanced Study; the Istituto Nazionale di Fisica Nucleare (INFN) sezione di Napoli, iniziative specifiche TEONGRAV; the International Max Planck Research School for Astronomy and Astrophysics at the Universities of Bonn and Cologne; DFG research grant "Jet physics on horizon scales and beyond" (grant No. 443220636); Joint Columbia/Flatiron Postdoctoral Fellowship (research at the Flatiron Institute is supported by the Simons Foundation); the Japan Ministry of Education, Culture, Sports, Science and Technology (MEXT; grant JPMXP1020200109); the Japan Society for the Promotion of Science (JSPS) Grant-in-Aid for JSPS Research Fellowship (JP17J08829); the Joint Institute for Computational Fundamental Science, Japan; the Key Research Program of Frontier Sciences, Chinese Academy of Sciences (CAS, grants QYZDJ-SSW-SLH057, QYZDJSSW-SYS008, ZDBS-LY-SLH011); the Leverhulme Trust Early Career Research Fellowship; the Max-Planck-Gesellschaft (MPG); the Max Planck Partner Group of the MPG and the CAS; the MEXT/JSPS KAKENHI (grants 18KK0090, JP21H01137, JP18H03721, JP18K13594, 18K03709, JP19K14761, 18H01245, 25120007, 23K03453,22H00157); the MICINN Research Projects PID2019-108995GB-C22, PID2022-140888NB-C22 ; the MIT International Science and Technology Initiatives (MISTI) Funds; the Ministry of Science and Technology (MOST) of Taiwan (103-2119-M-001-010-MY2, 105-2112-M-001-025-MY3, 105-2119-M-001-042, 106-2112-M-001-011, 106-2119-M-001-013, 106-2119-M-001-027, 106-2923-M-001-005, 107-2119-M-001-017, 107-2119-M-001-020, 107-2119-M-001-041, 107-2119-M-110-005, 107-2923-M-001-009, 108-2112-M-001-048, 108-2112-M-001-051, 108-2923-M-001-002, 109-2112-M-001-025, 109-2124-M-001-005, 109-2923-M-001-001, 110-2112-M-003-007-MY2, 110-2112-

M-001-033, 110-2124-M-001-007, and 110-2923-M-001-001); the Ministry of Education (MoE) of Taiwan Yushan Young Scholar Program; the Physics Division, National Center for Theoretical Sciences of Taiwan; the National Aeronautics and Space Administration (NASA, Fermi Guest Investigator grant 80NSSC20K1567, NASA Astrophysics Theory Program grant 80NSSC20K0527, NASA NuSTAR award 80NSSC20K0645); NASA Hubble Fellowship grants HST-HF2-51431.001-A, HST-HF2-51482.001-A, HST-HF2-51539.001-A awarded by the Space Telescope Science Institute, which is operated by the Association of Universities for Research in Astronomy, Inc., for NASA, under contract NAS5-26555; the National Institute of Natural Sciences (NINS) of Japan; the National Key Research and Development Program of China (grant 2016YFA0400704, 2017YFA0402703, 2016YFA0400702); the National Science and Technology Council (NSTC, grants NSTC 111-2112-M-001 -041, NSTC 111-2124-M-001-005, NSTC 112-2124-M-001-014); the US National Science Foundation (NSF, grants AST-0096454, AST-0352953, AST-0521233, AST-0705062, AST-0905844, AST-0922984, AST-1126433, OIA-1126433, AST-1140030, DGE-1144085, AST-1207704, AST-1207730, AST-1207752, MRI-1228509, OPP-1248097, AST-1310896, AST-1440254, AST-1555365, AST-1614868, AST-1615796, AST-1715061, AST-1716327, AST-1726637, OISE-1743747, AST-1743747, AST-1816420, AST-1952099, AST-1935980, AST-2034306, AST-2205908, AST-2307887); NSF Astronomy and Astrophysics Postdoctoral Fellowship (AST-1903847); the Natural Science Foundation of China (grants 11650110427, 10625314, 11721303, 11725312, 11873028, 11933007, 11991052, 11991053, 12192220, 12192223, 12273022, 12325302, 12303021); the Natural Sciences and Engineering Research Council of Canada (NSERC, including a Discovery Grant and the NSERC Alexander Graham Bell Canada Graduate Scholarships-Doctoral Program); the National Youth Thousand Talents Program of China; the National Research Foundation of Korea (the Global PhD Fellowship Grant: grants NRF-2015H1A2A1033752, the Korea Research Fellowship Program: NRF-2015H1D3A1066561, Brain Pool Program: RS-2024-00407499, Basic Research Support Grant 2019R1F1A1059721, 2021R1A6A3A01086420, 2022R1C1C1005255, 2022R1F1A1075115); Netherlands Research School for Astronomy (NOVA) Virtual Institute of Accretion (VIA) postdoctoral fellowships; NOIRLab, which is managed by the Association of Universities for Research in Astronomy (AURA) under a cooperative agreement with the National Science Foundation; Onsala Space Observatory (OSO) national infrastructure, for the provisioning of its facilities/observational support (OSO receives funding through the Swedish Research Council under grant 2017-00648); the Perimeter Institute for Theoretical Physics (research at Perimeter Institute is supported by the Government of Canada through the Department of Innovation, Science and Economic Development and by the Province of Ontario through the Ministry of Research, Innovation and Science); the Portuguese Foundation for Science and Technology (FCT) grants (Individual CEEC program - 5th edition, https://doi.org/10.54499/UIDB/04106/2020, https://doi.org/10.54499/UIDP/04106/2020, PTDC/FIS-AST/3041/2020, CERN/FIS-PAR/0024/2021, 2022.04560.PTDC); the Princeton Gravity Initiative; the Spanish Ministerio de Ciencia e Innovación (grants PGC2018-098915-B-C21, AYA2016-80889-P, PID2019-108995GB-C21, PID2020-117404GB-C21); the University of Pretoria for financial aid in the provision of the new Cluster Server nodes and SuperMicro (USA) for a SEEDING GRANT approved toward these nodes in 2020; the Shanghai Municipality orientation program of basic research for international scientists (grant no. 22JC1410600); the Shanghai Pilot Program for Basic Research, Chinese Academy of Science, Shanghai Branch (JCYJ-SHFY-2021-013); the State Agency for Research of the Spanish MCIU through the "Center of Excellence Severo Ochoa" award for the Instituto de Astrofísica de Andalucía (SEV-2017- 0709); the Spanish Ministry for Science and Innovation grant CEX2021-001131-S funded by MCIN/AEI/10.13039/501100011033; the Spinoza Prize SPI 78-409; the South African Research Chairs Initiative, through the South African Radio Astronomy Observatory (SARAO, grant ID 77948), which is a facility of the National Research Foundation (NRF), an agency of the Department of Science and Innovation (DSI) of South Africa; the Toray Science Foundation; the Swedish Research Council (VR); the UK Science and Technology Facilities Council (grant no. ST/X508329/1); the US Department of Energy (USDOE) through the Los Alamos National Laboratory (operated by Triad National Security, LLC, for the National Nuclear Security Administration of the USDOE, contract 89233218CNA000001); and the YCAA Prize Postdoctoral Fellowship; Conicyt through Fondecyt Postdoctorado (project 3220195).

We thank the staff at the participating observatories, correlation centers, and institutions for their enthusiastic support. This paper makes use of the following ALMA data: ADS/JAO.ALMA#2016.1.01290.V. ALMA is a partnership of the European Southern Observatory (ESO; Europe, representing its member states), NSF, and National Institutes of Natural Sciences of Japan, together with National Research Council (Canada), Ministry of Science and Technology (MOST; Taiwan), Academia Sinica Institute of Astronomy and Astrophysics (ASIAA; Taiwan), and Korea Astronomy and Space Science Institute (KASI; Republic of Korea), in cooperation with the Republic of Chile. The Joint ALMA Observatory is operated by ESO, Associated Universities, Inc. (AUI)/NRAO, and the National Astronomical Observatory of Japan (NAOJ). The NRAO is a facility of the NSF operated under cooperative agreement by AUI. This research used resources of the Oak Ridge Leadership Computing Facility at the Oak Ridge National Laboratory, which is supported by the Office of Science of the U.S. Department of Energy under contract No. DE-AC05-00OR22725; the ASTROVIVES FEDER infrastructure, with project code IDIFEDER-2021-086; the computing cluster of Shanghai VLBI correlator supported by the Special Fund for Astronomy from the Ministry of Finance in China; We also thank the Center for Computational Astrophysics, National Astronomical Observatory of Japan. This work was supported by FAPESP (Fundacao de Amparo a Pesquisa do Estado de Sao Paulo) under grant 2021/01183-8.

APEX is a collaboration between the Max-Planck-Institut für Radioastronomie (Germany), ESO, and the Onsala Space Observatory (Sweden). The SMA is a joint project between the SAO and ASIAA and is funded by the Smithsonian Institution and the Academia Sinica. The JCMT is operated by the East Asian Observatory on behalf of the NAOJ, ASIAA, and KASI, as well as the Ministry of Finance of China, Chinese Academy of Sciences, and the National Key Research and Development Program (No. 2017YFA0402700) of China and Natural Science Foundation of China grant 11873028. Additional funding support for the JCMT is provided by the Science and Technologies Facility Council (UK) and participating universities in the UK and Canada. The LMT is a project operated by the Instituto Nacional de Astrófisica, Óptica, y Electrónica (Mexico) and the University of Massachusetts at Amherst (USA). The IRAM 30-m telescope on Pico Veleta, Spain is operated by IRAM and supported by CNRS (Centre National de la Recherche Scientifique, France), MPG (Max-Planck-Gesellschaft, Germany), and IGN (Instituto Geográfico Nacional, Spain). The SMT is operated by the Arizona Radio Observatory, a part of the Steward Observatory of the University of Arizona, with financial support of operations from the State of Arizona and financial support for instrumentation development from the NSF. Support for SPT participation in the EHT is provided by the National Science Foundation through award OPP-1852617 to the University of Chicago. Partial support is also provided by the Kavli Institute of Cosmological Physics at the University of Chicago. The SPT hydrogen maser was provided on loan from the GLT, courtesy of ASIAA.

This work used the Extreme Science and Engineering Discovery Environment (XSEDE), supported by NSF grant ACI-1548562, and CyVerse, supported by NSF grants DBI-0735191, DBI-1265383, and DBI-1743442. XSEDE Stampede2 resource at TACC was allocated through TG-AST170024 and TG-AST080026N. XSEDE JetStream resource at PTI and TACC was allocated through AST170028. This research is part of the Frontera computing project at the Texas Advanced Computing Center through the Frontera Large-Scale Community Partnerships allocation AST20023. Frontera is made possible by National Science Foundation award OAC-1818253. This research was done using services provided by the OSG Consortium (Pordes et al. 2007; Sfiligoi et al. 2009) supported by the National Science Foundation award Nos. 2030508 and 1836650. Additional work used ABACUS2.0, which is part of the eScience center at Southern Denmark University, and the Kultrun Astronomy Hybrid Cluster (projects Conicyt Programa de Astronomia Fondo Quimal QUIMAL170001, Conicyt PIA ACT172033, Fondecyt Iniciacion 11170268, Quimal 220002). Simulations were also performed on the SuperMUC cluster at the LRZ in Garching, on the LOEWE cluster in CSC in Frankfurt, on the HazelHen cluster at the HLRS in Stuttgart, and on the Pi2.0 and Siyuan Mark-I at Shanghai Jiao Tong University. The computer resources of the Finnish IT Center for Science (CSC) and the Finnish Computing Competence Infrastructure (FCCI) project are acknowledged. This research was enabled in part by support provided by Compute Ontario (http://computeontario.ca), Calcul Quebec (http://www.calculquebec.ca), and Compute Canada (http://www.computecanada.ca).

The EHTC has received generous donations of FPGA chips from Xilinx Inc., under the Xilinx University Program. The EHTC has benefited from technology shared under open-source license by the Collaboration for Astronomy Signal Processing and Electronics Research (CASPER). The EHT project is grateful to T4Science and Microsemi for their assistance with hydrogen masers. This research has made use of NASA's Astrophysics Data System. We gratefully acknowledge the support provided by the extended staff of the ALMA, from the inception of the ALMA Phasing Project through the observational campaigns of 2017 and 2018. We would like to thank A. Deller and W. Brisken for EHT-specific support with the use of DiFX. We thank Martin Shepherd for the addition of extra features in the Difmap software that were used for the CLEAN imaging results presented in this paper. We acknowledge the significance that Maunakea, where the SMA and JCMT EHT stations are located, has for the indigenous Hawaiian people.

## Appendix B: Extended tables and figures

Here we provide additional figures and tables, which are of relevance for a full documentation of our data analysis.

Tables 1 list all parameters of identified model components for the VLBA observations at 1.6, 5.0, and 8.4 GHz and Tab. 2 provides a list of the model components for VLBA observations at 15, 22, and 43 GHz and for the 86 GHz GMVA observation. Figure 1 provides an example spectral index map between 22 and 43 GHz for the VLBA observations, Fig. 2 provides spectral fits for component B6 and A15 for an alternative interpretation of the central component at 22 GHz, in which Component B6 is identified with the EHT measurement and central GMVA component instead of A15. Fig. 3 shows the best fit obtained for all identified model components within the inner 20 mas of the jets. The corresponding fitting parameters are listed in Tab. 4. Figure 4 shows clean maps and model components for all six VLBA observations with the maps and components being shifted according to the 2D cross-correlation alignment.

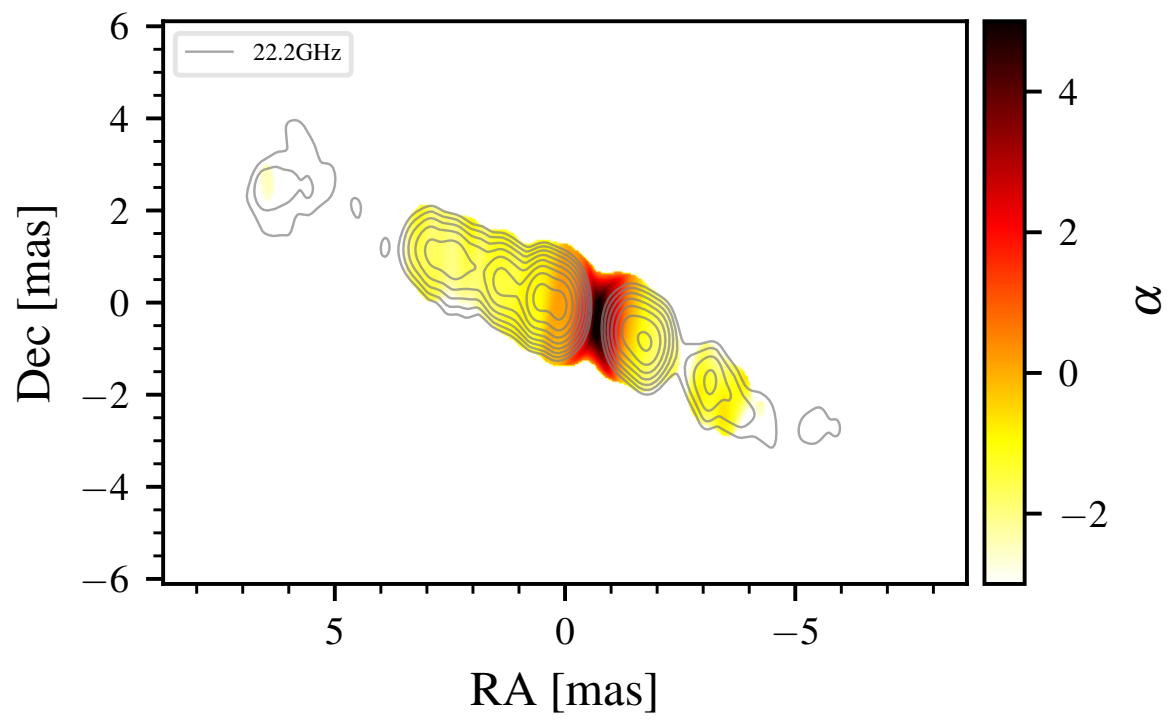

**Fig. 1:** Spectral index map between 22 and 43 GHz after alignment and convolving both maps with a mean, circular beam. The contours correspond to the 22 GHz image and starts at three times the noise level.

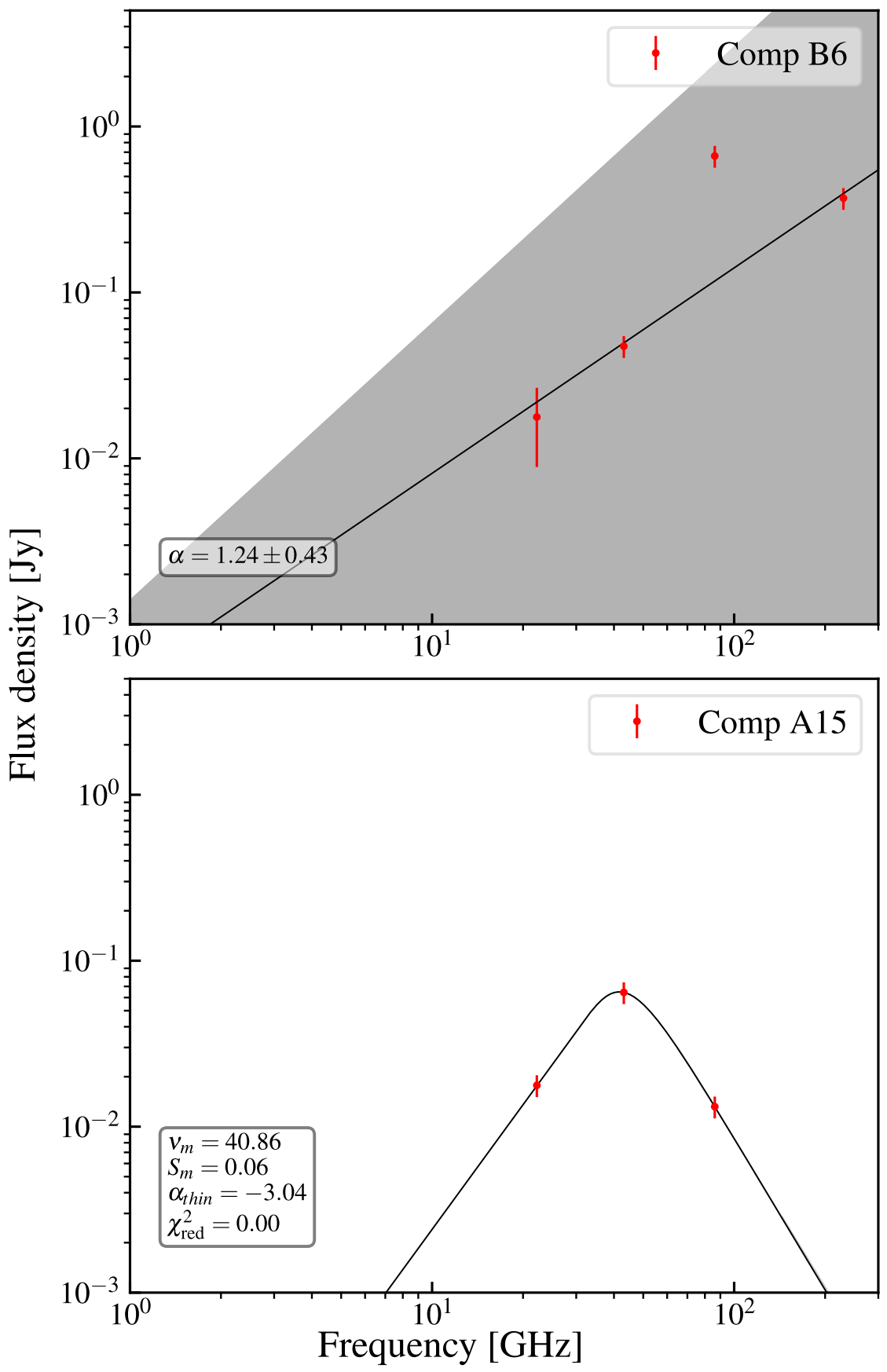

**Fig. 2:** Alternative continuum spectrum of components *A15* and *B6* if identifying the central 86 GHz component and the EHT data point with *B6*. For component *B6* the gray shaded area shows the 95% confidence interval of the power-law fit (black line). For *A15* a SSA-function with $\alpha_{\text{thick}} = 2.5$ was fitted to the data (black line), as only 3 points are provided no confidence interval is shown in this case.

**Table 1:** Parameters of Gaussian model fitting components for 1.6, 5.0, and 8.4 GHz VLBA observation for NGC 1052. Corresponding to shifted component positions.

| ID | Flux density [mJy] | RA [mas] | RA Error [mas] | DEC [mas] | DEC Error [mas] | Major [1] [mas] | Major Error [2] [mas] | Ratio [3] | Distance [mas] | Distance Error [4] [mas] | log $T_b$ [K] |
|---|---|---|---|---|---|---|---|---|---|---|---|
| | | | | | 1.6 GHz - Band | | | | | | |
| EJ | 10.9 | 57.54 | 1.29 | 20.14 | 0.54 | 15.92 | 1.86 | 1.00 | 61.42 | 0.55 | 7.34 |
| A1 | 7.6 | 33.61 | 1.29 | 15.28 | 0.54 | 6.17 | 1.86 | 1.00 | 37.18 | 0.55 | 8.01 |
| A2 | 34.1 | 27.12 | 1.29 | 12.66 | 0.54 | 5.18 | 1.86 | 1.00 | 30.18 | 0.55 | 8.81 |
| A3 | 22.9 | 22.83 | 1.29 | 10.24 | 0.54 | 2.36 | 1.86 | 1.00 | 25.33 | 0.55 | 9.32 |
| A4 | 38.2 | 19.24 | 1.29 | 9.65 | 0.54 | 1.86 | 1.86 | 1.00 | 21.75 | 0.55 | > 9.75 |
| A5 | 37.1 | 15.84 | 1.29 | 6.98 | 0.54 | 2.19 | 1.86 | 1.00 | 17.71 | 0.54 | 9.60 |
| A6 | 40.7 | 11.09 | 1.29 | 4.86 | 0.54 | 3.48 | 1.86 | 1.00 | 12.69 | 0.54 | 9.23 |
| A7 | 323.7 | 5.63 | 1.29 | 2.62 | 0.54 | 1.89 | 1.86 | 1.00 | 5.49 | 1.16 | 10.67 |
| A9/10 | 53.9 | 2.24 | 1.29 | 1.02 | 0.54 | 2.38 | 1.86 | 1.00 | 2.35 | 0.65 | 9.69 |
| D2 | 13.3 | −19.57 | 1.29 | −9.05 | 0.54 | 8.62 | 1.86 | 1.00 | −21.42 | 0.57 | 7.96 |
| D1 | 28.4 | −25.29 | 1.29 | −10.75 | 0.54 | 4.44 | 1.86 | 1.00 | −27.28 | 0.57 | 8.87 |
| | | | | | 5.0 GHz - Band | | | | | | |
| A2 | 7.9 | 26.83 | 0.38 | 12.90 | 0.17 | 5.39 | 0.55 | 1.00 | 29.79 | 0.17 | 7.13 |
| A3 | 2.2 | 22.93 | 0.38 | 11.24 | 0.17 | 0.55 | 0.55 | / | 25.56 | / | > 8.55 |
| A4 | 7.7 | 19.92 | 0.38 | 9.88 | 0.17 | 2.87 | 0.55 | 1.00 | 22.25 | 0.17 | 7.66 |
| A5 | 8.3 | 16.26 | 0.38 | 7.27 | 0.17 | 2.61 | 0.55 | 1.00 | 17.84 | 0.17 | 7.78 |
| A6 | 8.1 | 11.75 | 0.38 | 4.51 | 0.17 | 3.98 | 0.55 | 1.00 | 12.64 | 0.17 | 7.40 |
| A7 | 166.3 | 6.03 | 0.38 | 2.85 | 0.17 | 1.42 | 0.55 | 1.00 | 6.70 | 0.17 | 9.61 |
| A8 | 16.1 | 4.44 | 0.38 | 2.93 | 0.17 | 0.55 | 0.55 | 1.00 | 5.32 | 0.18 | > 9.41 |
| A9 | 176.1 | 2.97 | 0.38 | 1.41 | 0.17 | 0.67 | 0.55 | 1.00 | 3.35 | 0.17 | 10.28 |
| A10 | 303.9 | 2.16 | 0.38 | 1.16 | 0.17 | 0.78 | 0.55 | 1.00 | 2.56 | 0.21 | 10.40 |
| A11/12 | 255.7 | 1.20 | 0.38 | 0.64 | 0.17 | 0.55 | 0.55 | 1.00 | 1.35 | 0.19 | 10.61 |
| A13/A14 | 33.5 | 0.35 | 0.38 | 0.33 | 0.17 | 0.55 | 0.55 | / | 0.47 | / | > 9.73 |
| C2 | 30.1 | −4.71 | 0.38 | −2.27 | 0.17 | 1.12 | 0.55 | 1.00 | −5.22 | 0.18 | 9.07 |
| C1 | 19.3 | −5.76 | 0.38 | −2.57 | 0.17 | 1.31 | 0.55 | 1.00 | −6.29 | 0.17 | 8.75 |
| WJ-D | 1.5 | −15.04 | 0.38 | −6.67 | 0.17 | 0.55 | 0.55 | / | −16.43 | / | > 8.39 |
| D2 | 2.2 | −21.02 | 0.38 | −9.67 | 0.17 | 4.98 | 0.55 | 1.00 | −23.12 | 0.17 | 6.63 |
| D1 | 5.3 | −25.39 | 0.38 | −10.42 | 0.17 | 5.31 | 0.58 | 0.41 | −27.41 | 0.17 | 7.35 |
| | | | | | 8.4 GHz - Band | | | | | | |
| A4 | 7.1 | 19.33 | 0.23 | 9.42 | 0.10 | 5.29 | 0.33 | 1.00 | 21.51 | 0.11 | 6.65 |
| A5 | 5.2 | 14.97 | 0.23 | 6.11 | 0.10 | 3.92 | 0.33 | 1.00 | 16.18 | 0.10 | 6.76 |
| A7 | 104.8 | 6.10 | 0.23 | 2.67 | 0.10 | 1.38 | 0.33 | 1.00 | 6.67 | 0.10 | 8.98 |
| A8 | 17.8 | 4.25 | 0.23 | 2.14 | 0.10 | 1.35 | 0.33 | 1.00 | 4.76 | 0.10 | 8.23 |
| A9 | 166.7 | 2.94 | 0.23 | 1.22 | 0.10 | 0.50 | 0.33 | 1.00 | 3.20 | 0.10 | 10.05 |
| A10 | 218.1 | 2.27 | 0.23 | 0.96 | 0.10 | 0.57 | 0.33 | 1.00 | 2.48 | 0.10 | 10.07 |
| A11/12 | 224.6 | 1.42 | 0.23 | 0.60 | 0.10 | 0.36 | 0.33 | 1.00 | 1.57 | 0.10 | 10.47 |
| A13 | 285.3 | 0.70 | 0.23 | 0.21 | 0.10 | 0.47 | 0.33 | 1.00 | 0.67 | 0.14 | 10.34 |
| B2 | 32.3 | −1.85 | 0.23 | −0.75 | 0.10 | 0.40 | 0.33 | 1.00 | −1.98 | 0.10 | 9.54 |
| B1 | 14.0 | −2.95 | 0.23 | −1.50 | 0.10 | 0.96 | 0.33 | 1.00 | −3.31 | 0.11 | 8.42 |
| C2 | 19.1 | −4.15 | 0.23 | −2.40 | 0.10 | 0.36 | 0.33 | 1.00 | −4.79 | 0.11 | 9.41 |
| C1 | 31.9 | −5.23 | 0.23 | −2.66 | 0.10 | 1.25 | 0.33 | 1.00 | −5.86 | 0.11 | 8.54 |
| WJ-C | 2.5 | −7.17 | 0.23 | −3.72 | 0.10 | 0.34 | 0.33 | 1.00 | −8.08 | 0.11 | 8.58 |
| D1 | 3.3 | −24.67 | 0.23 | −12.10 | 0.10 | 2.84 | 0.33 | 1.00 | −27.47 | 0.11 | 6.85 |

[1] FWHM major axis of restoring beam [2] Error on Major ax equal to 1/5th beam [3] Ratio of minor to major axis
[4] Error on Position equal to 1/10th beam

**Table 2:** Parameters of Gaussian model fitting components for 15, 22, and 43 GHz VLBA for NGC 1052. Corresponding to shifted component positions.

| ID | Flux density [mJy] | RA [mas] | RA Error [mas] | DEC [mas] | DEC Error [mas] | Major [1] [mas] | Major Error [2] [mas] | Ratio [3] | Distance [mas] | Distance Error [4] [mas] | log $T_b$ [K] |
|---|---|---|---|---|---|---|---|---|---|---|---|
| 15.3 GHz - Band | | | | | | | | | | | |
| A7 | 37.4 | 6.05 | 0.12 | 2.67 | 0.06 | 1.40 | 0.18 | 1.00 | 6.69 | 0.06 | 8.00 |
| A8 | 5.3 | 4.56 | 0.12 | 2.68 | 0.06 | 0.67 | 0.18 | 1.00 | 5.33 | 0.06 | 7.79 |
| A9 | 70.9 | 2.91 | 0.12 | 1.28 | 0.06 | 0.63 | 0.18 | 1.00 | 3.27 | 0.06 | 8.97 |
| A10 | 78.7 | 2.32 | 0.12 | 0.98 | 0.06 | 0.60 | 0.18 | 1.00 | 2.62 | 0.06 | 9.06 |
| A11 | 62.8 | 1.52 | 0.12 | 0.73 | 0.06 | 0.33 | 0.18 | 1.00 | 1.78 | 0.05 | 9.49 |
| A12 | 60.4 | 1.23 | 0.12 | 0.49 | 0.06 | 0.24 | 0.18 | 1.00 | 1.49 | 0.06 | 9.73 |
| A13 | 274.9 | 0.58 | 0.12 | 0.26 | 0.06 | 0.36 | 0.18 | 1.00 | 1.03 | 0.11 | 10.05 |
| A14 | 73.5 | 0.12 | 0.12 | 0.02 | 0.06 | 0.24 | 0.18 | 1.00 | 0.09 | 0.08 | 9.83 |
| B3 | 28.4 | −1.46 | 0.12 | −0.41 | 0.06 | 0.22 | 0.18 | 1.00 | −1.44 | 0.06 | 9.47 |
| B2 | 102.3 | −1.80 | 0.12 | −0.75 | 0.06 | 0.35 | 0.18 | 1.00 | −1.90 | 0.06 | 9.64 |
| B1 | 19.1 | −3.05 | 0.12 | −1.58 | 0.06 | 0.85 | 0.18 | 1.00 | −3.39 | 0.06 | 8.15 |
| C2 | 11.8 | −3.93 | 0.12 | −2.23 | 0.06 | 0.71 | 0.18 | 1.00 | −4.48 | 0.06 | 8.09 |
| C1 | 13.8 | −5.16 | 0.12 | −2.64 | 0.06 | 1.44 | 0.18 | 1.00 | −5.75 | 0.06 | 7.54 |
| 22.2 GHz - Band | | | | | | | | | | | |
| A7 | 21.0 | 6.01 | 0.08 | 2.65 | 0.03 | 1.39 | 0.12 | 1.00 | 6.66 | 0.04 | 7.43 |
| A8 | 3.5 | 4.42 | 0.08 | 2.16 | 0.03 | 0.41 | 0.12 | 1.00 | 5.00 | 0.04 | 7.71 |
| A9 | 39.1 | 2.88 | 0.08 | 1.28 | 0.03 | 0.56 | 0.12 | 1.00 | 3.23 | 0.04 | 8.49 |
| A10 | 29.4 | 2.40 | 0.08 | 1.05 | 0.03 | 0.35 | 0.12 | 1.00 | 2.71 | 0.04 | 8.79 |
| A10/11 | 27.0 | 1.96 | 0.08 | 0.84 | 0.03 | 0.53 | 0.12 | 1.00 | 2.23 | 0.04 | 8.38 |
| A11 | 49.7 | 1.40 | 0.08 | 0.58 | 0.03 | 0.27 | 0.12 | 1.00 | 1.61 | 0.04 | 9.22 |
| A12 | 25.7 | 1.06 | 0.08 | 0.46 | 0.03 | 0.19 | 0.12 | 1.00 | 1.25 | 0.03 | 9.24 |
| A13 | 235.4 | 0.51 | 0.08 | 0.26 | 0.03 | 0.33 | 0.12 | 1.00 | 0.67 | 0.03 | 9.74 |
| A14 | 159.5 | 0.11 | 0.08 | 0.05 | 0.03 | 0.16 | 0.12 | 1.00 | 0.32 | 0.06 | 10.17 |
| A15 | 35.5 | −0.12 | 0.08 | −0.04 | 0.03 | 0.18 | 0.12 | 1.00 | −0.08 | 0.05 | 9.42 |
| B3 | 64.8 | −1.45 | 0.08 | −0.44 | 0.03 | 0.38 | 0.12 | 1.00 | −1.42 | 0.04 | 9.05 |
| B2 | 95.5 | −1.81 | 0.08 | −0.78 | 0.03 | 0.35 | 0.12 | 1.00 | −1.90 | 0.04 | 9.28 |
| B1 | 18.6 | −3.18 | 0.08 | −1.56 | 0.03 | 0.49 | 0.12 | 1.00 | −3.47 | 0.04 | 8.29 |
| C2 | 11.7 | −3.93 | 0.08 | −2.24 | 0.03 | 0.78 | 0.12 | 1.00 | −4.45 | 0.04 | 7.68 |
| C1 | 5.8 | −5.74 | 0.08 | −2.56 | 0.03 | 0.71 | 0.12 | 1.00 | −6.20 | 0.04 | 7.46 |
| 43.1 GHz - Band | | | | | | | | | | | |
| A7 | 2.2 | 6.29 | 0.04 | 2.60 | 0.02 | 0.66 | 0.06 | 1.00 | 6.81 | 0.02 | 6.52 |
| A9 | 16.5 | 2.91 | 0.04 | 1.27 | 0.02 | 0.58 | 0.06 | 1.00 | 3.17 | 0.02 | 7.52 |
| A10 | 2.5 | 2.38 | 0.04 | 1.25 | 0.02 | 0.11 | 0.06 | 1.00 | 2.69 | 0.02 | 8.12 |
| A10/11 | 15.9 | 2.06 | 0.04 | 0.97 | 0.02 | 0.53 | 0.06 | 1.00 | 2.28 | 0.02 | 7.58 |
| A11 | 16.5 | 1.44 | 0.04 | 0.73 | 0.02 | 0.25 | 0.06 | 1.00 | 1.61 | 0.02 | 8.24 |
| A12 | 24.0 | 1.01 | 0.04 | 0.35 | 0.02 | 0.42 | 0.06 | 1.00 | 1.07 | 0.02 | 7.94 |
| EJ | 23.8 | 0.72 | 0.04 | 0.25 | 0.02 | 0.07 | 0.06 | 1.00 | 0.76 | 0.02 | 9.46 |
| A13 | 85.5 | 0.50 | 0.04 | 0.32 | 0.02 | 0.19 | 0.06 | 1.00 | 0.59 | 0.02 | 9.19 |
| A14 | 161.7 | 0.17 | 0.04 | 0.13 | 0.02 | 0.12 | 0.06 | 1.00 | 0.17 | 0.04 | 9.88 |
| A15 | 64.4 | 0.00 | 0.04 | 0.01 | 0.02 | 0.13 | 0.06 | 1.00 | 0.00 | 0.02 | 9.38 |
| B6 | 47.4 | −0.23 | 0.04 | −0.05 | 0.02 | 0.11 | 0.06 | 1.00 | −0.23 | 0.02 | 9.40 |
| B5 | 18.8 | −0.73 | 0.04 | −0.14 | 0.02 | 0.34 | 0.06 | 1.00 | −0.74 | 0.02 | 8.03 |
| B4 | 39.4 | −1.09 | 0.04 | −0.39 | 0.02 | 0.16 | 0.06 | 1.00 | −1.15 | 0.02 | 8.99 |
| B3 | 37.2 | −1.33 | 0.04 | −0.39 | 0.02 | 0.24 | 0.06 | 1.00 | −1.38 | 0.02 | 8.62 |
| B2 | 50.3 | −1.68 | 0.04 | −0.67 | 0.02 | 0.36 | 0.06 | 1.00 | −1.81 | 0.02 | 8.42 |
| B1 | 9.2 | −2.91 | 0.04 | −1.58 | 0.02 | 0.44 | 0.06 | 1.00 | −3.31 | 0.02 | 7.50 |
| C2 | 4.9 | −3.73 | 0.04 | −2.07 | 0.02 | 0.25 | 0.06 | 1.00 | −4.27 | 0.02 | 7.70 |

[1] FWHM major axis of restoring beam [2] Error on Major ax equal to 1/5th beam [3] Ratio of minor to major axis
[4] Error on Position equal to 1/10th beam

**Table 3:** Parameters of Gaussian model fitting components for the 86 GHz GMVA observation for NGC 1052. Corresponding to shifted component positions.

| ID | Flux density [mJy] | RA [mas] | RA Error [mas] | DEC [mas] | DEC Error [mas] | Major [1] [mas] | Major Error [2] [mas] | Ratio [3] | Distance [mas] | Distance Error [4] [mas] | log $T_b$ [K] |
|---|---|---|---|---|---|---|---|---|---|---|---|
| 86.2 GHz - Band | | | | | | | | | | | |
| EJ | 59.3 | 0.79 | 0.03 | 0.23 | 0.01 | 0.25 | 0.04 | 1.00 | 0.82 | 0.01 | 8.19 |
| A13 | 18.5 | 0.51 | 0.03 | 0.27 | 0.01 | 0.04 | 0.04 | 1.00 | 0.58 | 0.01 | > 9.27 |
| A14 | 13.2 | 0.21 | 0.03 | 0.08 | 0.01 | 0.04 | 0.04 | 1.00 | 0.23 | 0.01 | > 9.13 |
| A15 | 663.8 | 0.00 | 0.03 | 0.01 | 0.01 | 0.04 | 0.04 | 1.00 | −0.01 | 0.02 | > 10.83 |
| B6 | 28.4 | −0.20 | 0.03 | −0.07 | 0.01 | 0.05 | 0.04 | 1.00 | −0.21 | 0.01 | 9.33 |
| B5 | 66.4 | −0.74 | 0.03 | −0.25 | 0.01 | 0.31 | 0.04 | 1.00 | −0.78 | 0.01 | 8.06 |

[1] FWHM major axis of restoring beam [2] Error on Major ax equal to 1/5th beam [3] Ratio of minor to major axis
[4] Error on Position equal to 1/10th beam

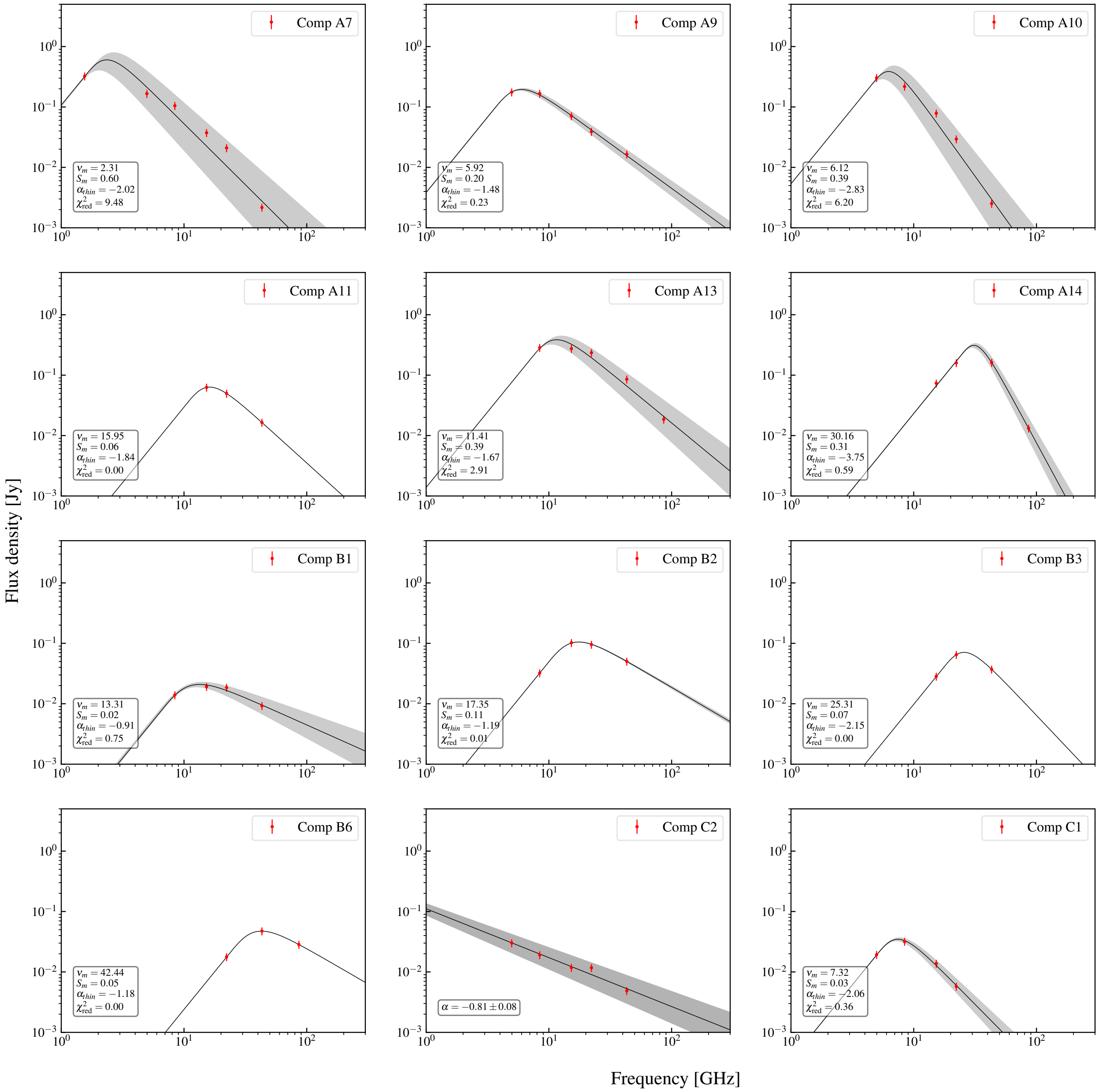

**Fig. 3:** Continuum spectrum of modelfit components within the inner 20 mas detected at at least 3 frequencies. For all components we attempted a simple power-law fit and a SSA-fit (with $\alpha_{\rm thick} = 2.5$). In the figures the fit (black line) with $\chi^2$ closer to 1 is plotted on top of the data points (red). In addition the gray shaded area corresponds to the 95% confidence interval (shown only for the cases with more than 3 data points).

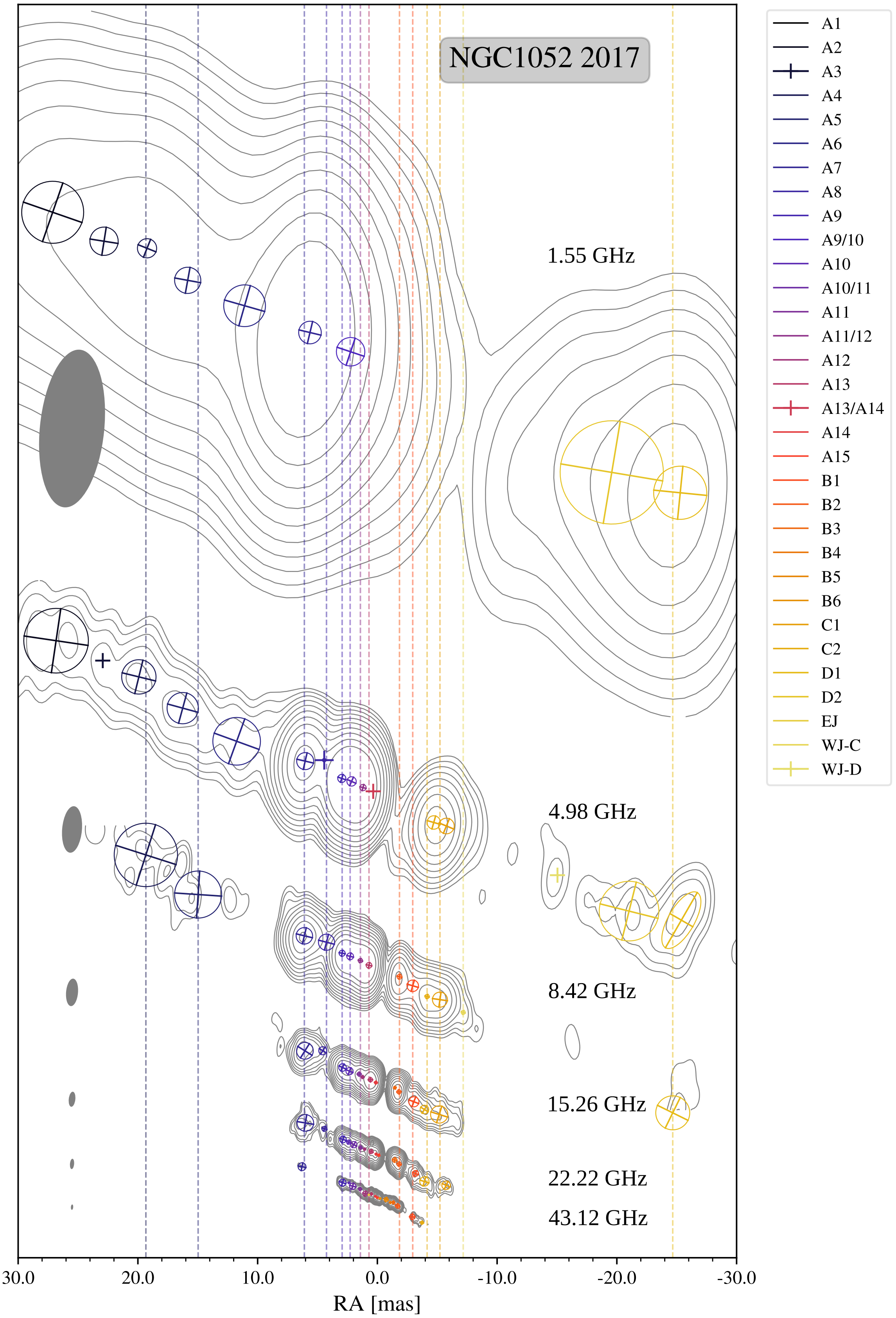

**Fig. 4:** Contour maps of images from 1.5 GHz to 43 GHz with Gaussian model components plotted on top. The DEC scale is equal to the RA scale. The VLBA maps have been shifted with respect to the 43 GHz image, based on the alignment described in **?**, by a 2D-cross-correlation. An additional shift was applied by identifying the Component A15 between the maps at 43 GHz and 86 GHz, and setting the component A15 at the map origin. The dashed lines correspond to the component positions at 5 GHz, the color corresponding to the respective component. The component names are assigned as 'A' for the eastern jet, 'B','C' and 'D' for the western jet from inside to outside.

**Table 4:** Fitting results of SSA spectrum to a representative model component selection. In the case of the same number of data points and free fit parameters we are not providing uncertainties of the fit parameters as these cannot be derived statistically correctly from our linear regression fit.

| ID | $\nu_m$[1] [GHz] | $S_m$[1] [Jy] | $\alpha_{thin}$[1] | $\chi^2$[1] |
|---|---|---|---|---|
| A7 | 2.31 ± 0.32 | 0.60 ± 0.20 | -2.02 ± 0.23 | 9.48 |
| A9 | 5.92 ± 0.25 | 0.20 ± 0.01 | -1.48 ± 0.07 | 0.23 |
| A10 | 6.12 ± 0.67 | 0.39 ± 0.10 | -2.83 ± 0.29 | 6.20 |
| A11 | 15.95 | 0.06 | -1.84 | - |
| A13 | 11.41 ± 1.09 | 0.39 ± 0.06 | -1.67 ± 0.19 | 2.91 |
| A14 | 30.16 ± 0.89 | 0.31 ± 0.03 | -3.75 ± 0.25 | 0.59 |
| A15[2] | 126.22 | 1.75 | -3.68 | - |
| B1 | 13.31 ± 0.78 | 0.02 ± 0.00 | -0.91 ± 0.19 | 0.75 |
| B2 | 17.35 ± 0.10 | 0.11 ± 0.00 | -1.19 ± 0.03 | 0.01 |
| B3 | 25.31 | 0.07 | -2.15 | - |
| B6 | 42.44 | 0.05 | -1.18 | - |
| C1 | 7.32 ± 0.23 | 0.03 ± 0.00 | -2.06 ± 0.16 | 0.36 |

[1] Fit parameters of SSA fit peak frequency, peak brightness, optically thin spectral index, and reduced $\chi^2$
[2] $\alpha_{thick} = 3.27$.